\documentclass[a4paper,fleqn]{cas-dc}

\usepackage[numbers,sort&compress]{natbib}

\usepackage{amsmath}
\usepackage{amssymb}
\usepackage{bm}
\usepackage{multirow}
\usepackage{subcaption}
\usepackage{booktabs}
\usepackage{tabularx}
\usepackage{placeins}

\usepackage[T1]{fontenc}

\usepackage{bm}
\usepackage{type1cm}                       
\usepackage{tikz}
\usetikzlibrary{arrows.meta}
\newcommand{\fscale}{0.95}   
\newcommand{\fs}[1]{\pgfmathsetmacro{\fsz}{#1*\fscale}\fontsize{\fsz pt}{\fsz pt}\selectfont}
\usetikzlibrary{calc}

\renewcommand{\_}{\textunderscore\allowbreak}

\newcommand{\panels}{\texttt{panels}}
\newcommand{\conn}[1]{\texttt{`#1'}}
\newcommand{\vect}[1]{\bm{#1}}
\newcommand{\mat}[1]{\bm{#1}}
\newcommand{\dd}{\,\mathrm{d}}

\begin{document}
\let\WriteBookmarks\relax

\shorttitle{Multidomain semi-analytical cohesive zone model for mode I delamination and damage progression}

\shortauthors{NTA D'Souza, BHAH Tijs, SGP Castro}

\title [mode = title]{Multidomain semi-analytical cohesive zone model for mode I delamination and damage progression}

\author[1]{Nathan T. A. D'Souza}[orcid=0009-0000-5384-6665]
\fnmark[1]
\credit{Methodology, Software, Formal analysis, Investigation,
Writing -- original draft}
\affiliation[1]{organization={Department of Aerospace Structures \& Materials,
            Faculty of Aerospace Engineering, Delft University of Technology},
            addressline={Kluyverweg 1},
            city={Delft},
            postcode={2629 HS},
            country={The Netherlands}}

\author[2]{Bas H. A. H. Tijs}[orcid=0000-0002-5506-1180]
\credit{Resources, Supervision, Validation, Writing -- review \& editing}
\affiliation[2]{organization={GKN Fokker Aerostructures},
            city={Papendrecht},
            country={The Netherlands}}

\author[1]{Saullo G. P. Castro}[orcid=0000-0001-9711-0991]
\cormark[1]
\fnmark[1]
\ead{S.G.P.Castro@tudelft.nl}
\credit{Conceptualisation, Supervision, Methodology, Software, Validation,
Writing -- original draft, Writing -- review \& editing}

\cortext[1]{Corresponding author}
\fntext[1]{These authors contributed equally to this work.}

\begin{abstract}
Semi-analytical Ritz models are much cheaper than finite element models for postbuckled stiffened panels, but rarely capture delamination initiation and growth. This work presents a multidomain semi-analytical cohesive zone model that tracks both along two-dimensional interfaces without prior knowledge of damage locations. Domains with independent Ritz approximations are connected through penalties, and the area connection between adjoining plates becomes a cohesive interface whose stiffness degrades at the integration points following a bilinear traction separation law. A consistent, non-symmetric tangent lets the Newton--Raphson solver follow the softening response past the peak, with a backtracking line search and increment bisection as safeguards. The interface stiffness is assembled exactly through matrix products, updated only at damaged integration points. Validation covers 10 mode I double cantilever beam cases from six references, including unidirectional and thermoplastic composites and conduction-welded joints. With the same cohesive law, peak load and the corresponding opening are within 0.5\% and 0.7\% of finite element results. Peak loads are within 2.1\% of corrected beam theory for all unidirectional specimens with nominal strengths, and 7.5--7.6\% below tests on conduction-welded joints, whose cohesive zone is long compared with the arm thickness. The model is a foundation for predicting skin--stiffener separation in such panels.

\end{abstract}


\begin{highlights}
\item Multidomain Ritz model with a cohesive interface tracks delamination onset and growth
\item Consistent tangent: about six Newton iterations per increment after the peak load
\item Exact matrix-product assembly of the interface, updated only at damaged points
\item Rules for Ritz terms and integration points derived from a 42-run convergence study
\item Validated on 10 DCB cases from six references, including conduction-welded joints
\end{highlights}


\maketitle

\section{Introduction}\label{sec:intro}

Composite stiffened panels are increasingly designed to operate beyond their buckling load, which requires the failure modes of the postbuckled regime to be predicted~\citep{Kassapoglou_DACS1_Book}. With the trend towards fastener-free joints, co-cured, bonded or welded, one of these failure modes is the separation of the skin from the stiffener~\citep{vanDooren_TP_SSS,skin_stiff_fig_bas}. Out-of-plane stresses develop at the skin--stiffener interface where the load is transferred between members of different stiffness, near the free edge of the flange, and in the postbuckled regime, where the half-waves of the skin peel it away from the stiffener~\citep{Kassapoglou_DACS1_Book}. Once formed, a disbond can change the critical buckling mode from half-waves on each side of the stiffener to a single half-wave centred on the disbond, with an abrupt reduction of the buckling load~\citep{Saullo_MD_Delam,Silva_Saullo_Beam_Delam}. Predicting where such damage initiates and how it grows is crucial to designing against it, and the models used must be cheap enough for the many analyses involved in the design process.

Kassapoglou and Lagace~\citep{Kassapoglou_Interlaminar_Stress_Uniax} computed interlaminar stresses with assumed stress fields that satisfy equilibrium and minimise the complementary energy. The approach was extended to the skin--stiffener interface under shear~\citep{Kassapoglou_SS_Stress_Shear}, where the stresses predict the onset of delamination together with a stress criterion, such as the quadratic delamination criterion~\citep{Brewer_QDC}. Robeson~\citep{Robeson_SS_Cyl_Shells_Papers,Robeson_PhD_Thesis} determined the interface stresses of stiffened plates and shallow shells with a Rayleigh--Ritz approach. For the double cantilever beam (DCB), beams on an elastic foundation with an embedded cohesive law give closed-form solutions: Cabello et al.~\citep{Cabello_Turon_Analytical_DCB} for bonded joints, \v{S}kec et al.~\citep{Skec2019IJF} for Timoshenko arms with a bilinear law through the three phases of the response, with an enhanced simple beam theory for data reduction~\citep{Skec2019CompB}, and Tian et al.~\citep{Tian2024} for shear-deformable arms; for mixed-mode I--II delamination, where no closed form is available, de Morais~\citep{deMorais2014} coupled cohesive zone beam models with Timoshenko beam finite elements. K\"{o}llner and V\"{o}llmecke~\citep{Kollner2018} and K\"{o}llner~\citep{Kollner2021} modelled the postbuckling of delaminated plates and the growth of the delamination with Rayleigh--Ritz formulations in generalised coordinates, with an energy release rate criterion. Recently, Butler et al.~\citep{Butler2026} proposed a strip-based method for stiffener debonding in postbuckled panels, in which the energy release rate follows from the sensitivity of the buckling loads to the bonded width; its debonding loads are within 1\% of a cohesive FE model for a co-cured flange and within 2\% of the tests and of a virtual crack closure FE model for a welded hat-stiffened panel, at a cost several orders of magnitude lower. These methods are efficient but restricted to specific geometries and loads, and those that predict growth need a pre-existing crack.

Cohesive zone models, which originate in the model of Hillerborg et al.~\citep{Hillerborg1976}, where stresses act across a crack as long as it is narrowly opened, relate the tractions of an interface to the displacement jump across it. With a bilinear law, the traction grows linearly up to the interface strength and then softens linearly to zero, and the area under the law equals the fracture toughness, so that initiation and propagation are predicted without an initial crack. In finite element (FE) models they are implemented as interface or decohesion elements~\citep{Schellekens1993,Alfano2001,Camanho2003,Turon_2006}. Alfano and Crisfield~\citep{Alfano2001} discussed the energetic equivalence with linear elastic fracture mechanics (LEFM), the spurious oscillations of the response for meshes that are coarse compared with the cohesive zone and the numerical issues of the solution.
The length of the cohesive zone~\citep{Rice1980,YangCox2005,HarperHallett2008} limits the element size. Turon et al.~\citep{Turon2007} relaxed this limit by lowering the interface strength at constant fracture toughness, which lengthens the cohesive zone over more elements, at the cost of less accurate stresses near the crack tip, and gave a closed-form choice of the penalty stiffness. Harper and Hallett~\citep{HarperHallett2008} showed that the reduction has limits: once a crack has initiated the response is insensitive to the strength, but a long cohesive zone softens the response before propagation and delays it, and comparable reductions of the mode II strength fail to predict the peak load. The strength therefore cannot be lowered freely where the onset of damage without a pre-existing crack is to be predicted, as at the skin--stiffener interface.
Multilinear laws represent the R-curve due to fibre bridging~\citep{Bas_charaterization,Abdel_Bas_Turon_MM_MLCL}, and mixed-mode laws the interaction of modes~\citep{Turon_2006,Turon_MM_CZ_1st}. Several recent developments aim at reducing the cost: extended FE plates and shells with cohesive delamination~\citep{Yazdani_XFEM_MMCZ_2015,Yazdani_XFEM_MMCZ_2017}, the scaled boundary finite element method with cohesive interfaces~\citep{Garg2020}, refined structural models of the Carrera unified formulation with cohesive zones~\citep{Kaleel2020,Mohammed2025}, and cohesive elements that overcome the cohesive zone limit on the element size~\citep{TostiBalducci2024,Ai2025}. The benchmarks of Krueger~\citep{Krueger2008,Krueger2012,Krueger2015} were developed to assess the delamination propagation capabilities of commercial FE codes.

Semi-analytical methods, where the displacement field of each domain is approximated by continuous functions in the sense of Ritz, lie between the analytical and the FE methods. Their advantage fades when a large or irregular structure is mapped as one domain, or when the solution localises~\citep{Campagna2024}, which motivated multidomain approaches that divide the structure into domains with their own approximation functions~\citep{Saullo_MD_Delam,Campagna2024}. Castro and Donadon~\citep{Saullo_MD_Delam} assembled Ritz domains with penalty connections whose stiffness follows from the laminate properties, with Bardell's hierarchical functions~\citep{Bardell_Shape_Ftns}, to study the effect of debonding defects on the buckling and vibration of stiffened panels; da Silva et al.~\citep{Silva_Saullo_Beam_Delam} studied shear buckling with a Rayleigh--Ritz model whose domains are joined with $C^1$ continuity instead of penalties. Naganarayana and Atluri~\citep{Naganarayana_energy,Naganarayana_Delam_Growth} computed the energy release rate along the front of delaminations in multi-plate FE models, and Fagiano et al.~\citep{CK_Domain_decom_Interface_Stress} recovered interlaminar stresses with an FE domain decomposition (FETI). Barbieri and Meo~\citep{Barbieri_Delam_2009,Barbieri_Delam_2011} used two-dimensional meshfree bi-domain models with a cohesive law, as a penalty along the crack line, and Hesar et al.~\citep{Hesar_Mode1_Delam,Pekbey_Mode2_Delam} two-dimensional meshfree models with cohesive segments. Milazzo and Oliveri modelled cracked multilayered plates~\citep{MilazzoOliveri2015} and stiffened panels with cracks and delaminations of fixed size, a delamination being two overlapping plate elements connected at its edges by penalties, with contact against interpenetration~\citep{Milazzo_MD_Delam}, with Ritz approaches, the postbuckling of variable angle tow stiffened panels~\citep{OliveriMilazzo2018}, and, with Gulizzi~\citep{Gulizzi_Oliveri_MD_Crack}, enriched the approximation at crack tips (X-Ritz). Ritz formulations were recently extended to progressive intralaminar damage, with continuum damage mechanics at the ply level~\citep{Campagna2023}; where the damage localises, a single-domain basis produces Gibbs oscillations, which an adaptive multidomain discretisation removes~\citep{Campagna2024}, and cohesive interlaminar laws were left for future work~\citep{Campagna2023}. Plate decomposition~\citep{Ghannadpour_CracK_MD_Boundary} and hierarchical Ritz sets~\citep{Oliveria_Crack_Ritz} were applied to cracked plates, and Vescovini and Bisagni~\citep{VescoviniBisagni2013} proposed a two-step Ritz procedure for the fast postbuckling analysis of stiffened panels; for the prediction of skin--stringer debonding, Vescovini et al.~\citep{VescoviniDavilaBisagni2013} resorted to simplified FE models.

For delamination, most of the semi-analytical and multidomain methods above either rely on a pre-existing crack and an LEFM criterion at the crack front, or predict only the onset of delamination, or study the effect of a disbond of fixed size, and the few Ritz models of progressive damage are limited to intralaminar damage~\citep{Campagna2023,Campagna2024}. The cohesive models of Barbieri and Meo \cite{Barbieri_Delam_2009, Barbieri_Delam_2011} and of Hesar et al. \cite{Hesar_Mode1_Delam} are one-dimensional interfaces with a point-like front, and the latter depend strongly on the number of nodes: a 10\% increase of the number of nodes increased the critical load by 45\%~\citep{Hesar_Mode1_Delam}. A cohesive zone in the area connection between two Ritz domains, whose penalty stiffness is degraded by a traction--separation law at the integration points, would track initiation and propagation over a two-dimensional interface without prior knowledge of the damage location.
To the authors' knowledge, no such multidomain semi-analytical cohesive zone model, with a verified formulation, a validated response and discretisation rules, is available.

The contributions of this work are:
\begin{enumerate}[(i)]
\item a complete formulation of the multidomain cohesive zone model;
\item the consistent, non-symmetric tangent stiffness of the damaged interface;
\item an exact assembly of the interface stiffness by matrix products, with an update restricted to the damaged points, whose cost grows with the size of the process zone and of the cracked region and not with the size of the interface;
\item a robust solution procedure with a linear predictor, a backtracking line search and the bisection of increments;
\item discretisation rules for the cohesive domain from a convergence study;
\item the verification of the implementation, and its validation against six references, with 10 mode~I DCB cases and measured initial compliances, FE and experimental results, including thermoplastic and conduction-welded joints;
\item an open-source implementation with its tests, and a dataset with the scripts and the data needed to reproduce every result.
\end{enumerate}
Section~\ref{sec:methodology} presents the formulation and the verification and validation strategies, Section~\ref{sec:results} the results and the limitations identified, and Section~\ref{sec:conclusion} the conclusions.
Appendix~\ref{app:corrections} documents the origin of the model and the corrections and improvements with respect to its earlier formulation and implementation, and Appendix~\ref{app:repro} presents the full data reproducibility map of this work.
\section{Methodology}\label{sec:methodology}

The methods presented herein are implemented in the open-source library \panels{}~\citep{panels2026} and the results of this work were obtained with its version 0.14.0 (Appendix~\ref{app:repro}).

\subsection{Formulation}\label{sec:formulation}

The main symbols are listed in Table~\ref{tab:nomenclature}. Vectors and matrices are written in bold; units are N, mm and MPa.

\begin{table*}[!tbp]
\caption{Main symbols.}\label{tab:nomenclature}
\footnotesize
\resizebox{\linewidth}{!}{%
\begin{tabular}{@{}p{0.11\textwidth}p{0.37\textwidth}p{0.11\textwidth}p{0.37\textwidth}@{}}
\toprule
Symbol & Definition & Symbol & Definition \\
\midrule
$a_p$, $b_p$ & dimensions of domain $p$ along $x$, $y$ & $k_o$ & initial (penalty) stiffness of the cohesive law \\
$\xi$, $\eta$ & natural coordinates of a domain & $\tau_o$ & interface strength \\
$f_i$, $g_j$ & approximation functions along $\xi$, $\eta$ & $G_{Ic}$ & mode~I fracture toughness \\
$m$, $n$ & number of terms along $x$, $y$ & $\Delta_o$, $\Delta_f$ & separations at damage onset and at failure \\
$m_{tsl}$, $n_{tsl}$ & terms of the cohesive domain along $x$, $y$ & $d$, $\hat d$ & damage variable, damage of the current separation \\
$\vect{c}$ & Ritz constants of the assembly & $d^h$ & damage history (last converged state) \\
$u$, $v$, $w$ & mid-plane displacements & $\chi$ & loading mask of the consistent tangent; in Eqs.~\eqref{eq:lefm}--\eqref{eq:chi}, crack length correction of beam theory \\
$\mat{A}$, $\mat{B}$, $\mat{D}$ & laminate stiffness matrices; $D_{11}$ bending stiffness & $\mat{K}_C$, $\mat{K}_G$ & constitutive and geometric tangent of the domains \\
$h$ & laminate thickness & $\mat{K}^{pen}_{conn}$ & penalty stiffness of the edge and perfect-bond connections \\
$d^t$, $d^b$ & distance from the mid-plane of the top and bottom domains to the interface & $\mat{K}^{coh}_s$ & secant stiffness of the cohesive interface \\
$\Delta_u$, $\Delta_v$, $\Delta_w$ & tangential and normal separations & $\mat{K}^{coh}_{\dot d}$ & damage-rate part of the consistent tangent \\
$\bar\Delta_w$ & separation that drives the damage & $\mat{K}^{pen}_{PD}$ & stiffness of the prescribed displacement penalty \\
$\mat{B}_\alpha$ & operator of the separation $\alpha = u, v, w$ & $\vect{R}$ & residual \\
$k_t$, $k_r$ & translational and rotational penalties of an edge connection & $\vect{f}^{pan}_{int}$ & internal force of the domains \\
$k_w$ & penalty of the prescribed displacement & $P$ & reaction of the prescribed displacement (load) \\
$w_p$ & prescribed displacement & $\delta$ & opening of the DCB at the loaded end \\
$n_x$, $n_y$ & Gauss--Legendre points of the cohesive domain & $a$, $a_0$ & crack length, initial crack length \\
$w_g$ & integration weight of point $g$ & $L$, $L_{tsl}$ & length of the specimen, of the cohesive domain \\
$\lambda$ & decay rate of the opening ahead of the process zone & $l_{cz}$ & cohesive zone length \\
\bottomrule
\end{tabular}}
\end{table*}

\subsubsection{Ritz approximation and kinematics of the domains}\label{sec:ritz}

Each domain $p$ is a flat laminated plate of dimensions $a_p \times b_p$, with natural coordinates $\xi = 2x/a_p - 1$ and $\eta = 2y/b_p - 1$. Following Castro and Donadon~\citep{Saullo_MD_Delam}, the mid-plane displacements are approximated as
\begin{equation}\label{eq:ritz}
u = \sum_{i=1}^{m}\sum_{j=1}^{n} f^u_i(\xi)\,g^u_j(\eta)\,c^u_{ij}, \quad
\bar{\vect{u}} = \begin{Bmatrix} u & v & w \end{Bmatrix}^\top = \mat{S}\vect{c},
\end{equation}
and similarly for $v$ and $w$, where $\mat{S}$ collects the products $f_i g_j$ and $\vect{c}$ the Ritz constants. The approximation functions are the hierarchical functions of Bardell~\citep{Bardell_Shape_Ftns}. For anisotropic plates, such Legendre-based polynomials converge faster than trigonometric functions and, when the integrals are evaluated in closed form, give sparser matrices~\citep{Vescovini_Legendre_better}. The first four are the cubic Hermite polynomials
\begin{equation}\label{eq:bardell1}
\begin{aligned}
f_1 &= \left(\tfrac{1}{2} - \tfrac{3}{4}\xi + \tfrac{1}{4}\xi^3\right) t_L, \\
f_2 &= \left(\tfrac{1}{8} - \tfrac{1}{8}\xi - \tfrac{1}{8}\xi^2 + \tfrac{1}{8}\xi^3\right) r_L, \\
f_3 &= \left(\tfrac{1}{2} + \tfrac{3}{4}\xi - \tfrac{1}{4}\xi^3\right) t_R, \\
f_4 &= \left(-\tfrac{1}{8} - \tfrac{1}{8}\xi + \tfrac{1}{8}\xi^2 + \tfrac{1}{8}\xi^3\right) r_R,
\end{aligned}
\end{equation}
where the flags $t_L, r_L, t_R, r_R \in \{0, 1\}$ enable the translation and the rotation of the left ($\xi = -1$) and right ($\xi = 1$) edges, and the higher-order terms, which vanish with their first derivative at both edges, are
\begin{equation}\label{eq:bardell2}
f_{i > 4}(\xi) = \sum_{q=0}^{\lfloor i/2 \rfloor} \frac{(-1)^q (2i - 2q - 7)!!}{2^q\, q!\, (i - 2q - 1)!}\, \xi^{i - 2q - 1},
\end{equation}
with $0!! = (-1)!! = 1$. The functions $g_j(\eta)$ are defined in the same way.

The domains follow the classical laminated plate theory (CLPT) with the nonlinear strains of von K\'{a}rm\'{a}n:
\begin{equation}\label{eq:strains}
\begin{aligned}
\vect{\varepsilon}^0 &= \begin{Bmatrix} u_{,x} + \frac{1}{2} w_{,x}^2 \\ v_{,y} + \frac{1}{2} w_{,y}^2 \\ u_{,y} + v_{,x} + w_{,x} w_{,y} \end{Bmatrix}, \quad
\vect{\kappa} = \begin{Bmatrix} -w_{,xx} \\ -w_{,yy} \\ -2 w_{,xy} \end{Bmatrix}, \\[4pt]
\begin{Bmatrix} \vect{N} \\ \vect{M} \end{Bmatrix} &= \begin{bmatrix} \mat{A} & \mat{B} \\ \mat{B} & \mat{D} \end{bmatrix} \begin{Bmatrix} \vect{\varepsilon}^0 \\ \vect{\kappa} \end{Bmatrix}
= \mat{F} \begin{Bmatrix} \vect{\varepsilon}^0 \\ \vect{\kappa} \end{Bmatrix}.
\end{aligned}
\end{equation}
Writing the generalised strains as $\vect{\varepsilon} = \mat{B}_{\varepsilon}\vect{c}$ and their variation as $\delta\vect{\varepsilon} = \bar{\mat{B}}\delta\vect{c}$, the internal force of the domains and its derivative are
\begin{equation}\label{eq:fint-pan}
\vect{f}^{pan}_{int} = \int_A \bar{\mat{B}}^\top \begin{Bmatrix} \vect{N} \\ \vect{M} \end{Bmatrix} \dd A, \qquad
\frac{\partial \vect{f}^{pan}_{int}}{\partial \vect{c}} = \mat{K}_C + \mat{K}_G,
\end{equation}
where $\mat{K}_C = \int_A \bar{\mat{B}}^\top \mat{F} \bar{\mat{B}} \dd A$ is the constitutive stiffness, with $\mat{F}$ the laminate matrix of Eq.~\eqref{eq:strains}, and $\mat{K}_G$ the geometric stiffness due to the membrane forces~\citep{Saullo_PhD}. The matrices of each domain are placed on the diagonal of those of the assembly.

\subsubsection{Connections and prescribed displacements}\label{sec:connections}

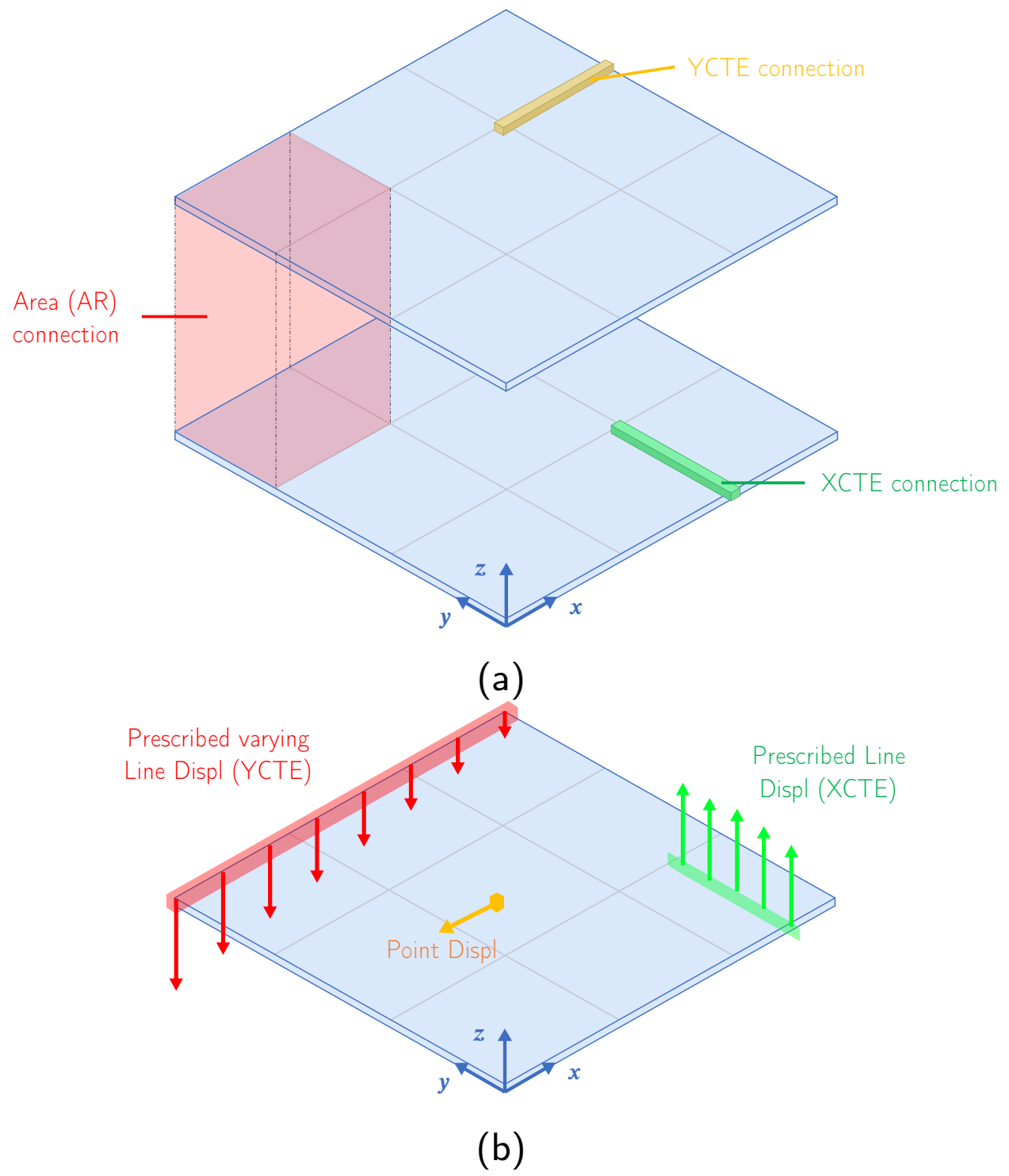
\begin{figure}
\centering
\begin{subfigure}{\columnwidth}
\resizebox{\columnwidth}{!}{%
\begin{tikzpicture}[x=1pt,y=-1pt,                 
  line cap=butt, line join=round,
  lab/.style={anchor=base,inner sep=0pt,outer sep=0pt,text height=0pt,text depth=0pt}]

\definecolor{plateFill}{RGB}{218,232,252}
\definecolor{plateEdge}{RGB}{62,110,194}
\definecolor{meshGrey}{RGB}{196,201,211}
\definecolor{axisBlue}{RGB}{62,110,194}
\definecolor{yteText}{RGB}{255,191,0}
\definecolor{xteText}{RGB}{0,255,4}
\definecolor{yteLine}{RGB}{255,87,0}
\definecolor{xteLine}{RGB}{0,176,80}
\definecolor{yelTop}{RGB}{233,216,151}\definecolor{yelSide}{RGB}{214,194,118}\definecolor{yelEnd}{RGB}{224,205,134}\definecolor{yelEdge}{RGB}{190,168,92}
\definecolor{grnTop}{RGB}{131,241,169}\definecolor{grnSide}{RGB}{96,214,138}\definecolor{grnEnd}{RGB}{112,226,152}\definecolor{grnEdge}{RGB}{70,185,110}
\useasboundingbox (-4,-4) rectangle (1336,832);   

\newcommand{\plate}[4]{%
  \fill[plateFill] (#1) -- (#4) -- ($(#4)+(0,10.5)$) -- ($(#1)+(0,10.5)$) -- cycle;
  \fill[plateFill] (#4) -- (#3) -- ($(#3)+(0,10.5)$) -- ($(#4)+(0,10.5)$) -- cycle;
  \fill[plateFill] (#1) -- (#2) -- (#3) -- (#4) -- cycle;
  \foreach \f in {0.346,0.691} \draw[meshGrey,line width=1.8pt] ($(#1)!\f!(#2)$) -- ($(#4)!\f!(#3)$);   
  \foreach \f in {0.303,0.655} \draw[meshGrey,line width=1.8pt] ($(#1)!\f!(#4)$) -- ($(#2)!\f!(#3)$);   
  }

\coordinate (L2) at (230.5,564.2); \coordinate (T2) at (672.8,314.8); \coordinate (R2) at (1115.0,564.2); \coordinate (B2) at (672.8,813.6);
\coordinate (L1) at (230.5,250.2); \coordinate (T1) at (672.5,0.9); \coordinate (R1) at (1114.5,250.2); \coordinate (B1) at (672.5,499.5);
\plate{L2}{T2}{R2}{B2}      
\plate{L1}{T1}{R1}{B1}      

\fill[red,opacity=0.2] (230.5,252) -- (384,165) -- (518,239) -- (518,554) -- (365.3,639) -- (230.5,563) -- cycle;

\begin{scope}[plateEdge,line width=1.5pt,line cap=round]
\draw[plateEdge] (230.5,250.2) -- (672.5,0.9);
\draw[plateEdge] (672.5,0.9) -- (1114.5,250.2);
\draw[plateEdge] (1114.5,250.2) -- (672.5,499.5);
\draw[plateEdge] (672.5,499.5) -- (230.5,250.2);
\draw[plateEdge] (230.5,260.7) -- (672.5,510.0);
\draw[plateEdge] (672.5,510.0) -- (1114.5,260.7);
\draw[plateEdge] (230.5,564.2) -- (672.8,813.6);
\draw[plateEdge] (672.8,813.6) -- (1115.0,564.2);
\draw[plateEdge] (230.5,564.2) -- (499.6,412.5);
\draw[plateEdge] (845.7,412.3) -- (1115.0,564.2);
\draw[plateEdge] (230.5,574.7) -- (672.8,824.1);
\draw[plateEdge] (672.8,824.1) -- (1115.0,574.7);
\draw[plateEdge] (230.5,250.2) -- (230.5,260.7);
\draw[plateEdge] (672.5,499.5) -- (672.5,510.0);
\draw[plateEdge] (1114.5,250.2) -- (1114.5,260.7);
\draw[plateEdge] (230.5,564.2) -- (230.5,574.7);
\draw[plateEdge] (672.8,813.6) -- (672.8,824.1);
\draw[plateEdge] (1115.0,564.2) -- (1115.0,574.7);
\end{scope}

\begin{scope}[black,opacity=0.6,line width=1.1pt,dash pattern=on 5pt off 2.5pt on 1.5pt off 2pt]
\draw (230.5,252)  -- (230.5,563);
\draw (365.3,325)  -- (365.3,639);
\draw (384,165)  -- (384,481.5);
\draw (518,239)    -- (518,554);
\end{scope}

\filldraw[fill=yelSide,draw=yelEdge,line width=0.6pt] (669.0,158.2) -- (816.0,72.9) -- (816.0,82.9) -- (669.0,168.2) -- cycle;   
\filldraw[fill=yelEnd,  draw=yelEdge,line width=0.6pt] (657.0,151.4) -- (669.0,158.2) -- (669.0,168.7) -- (657.0,161.9) -- cycle;   
\filldraw[fill=yelTop,  draw=yelEdge,line width=0.6pt] (657.0,151.4) -- (805.0,67.8) -- (816.0,72.9) -- (669.0,158.2) -- cycle;           
\filldraw[fill=grnSide,draw=grnEdge,line width=0.6pt] (813.0,555.8) -- (973.0,646.0) -- (973.0,656.5) -- (813.0,566.3) -- cycle;   
\filldraw[fill=grnEnd,  draw=grnEdge,line width=0.6pt] (985.0,639.2) -- (973.0,646.0) -- (973.0,656.5) -- (985.0,649.7) -- cycle;   
\filldraw[fill=grnTop,  draw=grnEdge,line width=0.6pt] (825.0,549.0) -- (985.0,639.2) -- (973.0,646.0) -- (813.0,555.8) -- cycle;           

\draw[red,line width=4pt]      (186,410.5) -- (273.5,410.5);
\draw[yteText,line width=4pt] (785,94.3) -- (898,77);
\draw[xteLine,line width=4pt]  (960,632.5) -- (1071.5,632.5);

\draw[axisBlue,line width=4.6pt,-{Triangle[length=17pt,width=18pt]}] (672.8,824.1) -- (672.8,738.8);  
\draw[axisBlue,line width=4.6pt,-{Triangle[length=17pt,width=18pt]}] (672.8,824.1) -- (741,784.5);    
\draw[axisBlue,line width=4.6pt,-{Triangle[length=17pt,width=18pt]}] (672.8,824.1) -- (605,785);    

\node[lab,text=red] at (84,401.1) {\fs{35.8}Area (AR)};
\node[lab,text=red] at (85,444.7) {\fs{35.6}connection};
\node[lab,text=yteText] at (1034.0,89.2) {\fs{35.8}YCTE connection};
\node[lab,text=xteLine] at (1212.0,644.3) {\fs{35.9}XCTE connection};
\node[lab,text=axisBlue] at (638,755.6) {\fs{34}$\bm{z}$};
\node[lab,text=axisBlue] at (766,807.6) {\fs{34}$\bm{x}$};
\node[lab,text=axisBlue] at (591,820.3) {\fs{34}$\bm{y}$};
\end{tikzpicture}}
\caption{}
\end{subfigure}
\begin{subfigure}{\columnwidth}
\resizebox{\columnwidth}{!}{%
\begin{tikzpicture}[x=1pt,y=-1pt,                 
  line cap=butt, line join=round,
  lab/.style={anchor=base,inner sep=0pt,outer sep=0pt,text height=0pt,text depth=0pt}]

\definecolor{plateFill}{RGB}{218,232,252}
\definecolor{plateEdge}{RGB}{62,110,194}
\definecolor{meshGrey}{RGB}{196,201,211}
\definecolor{axisBlue}{RGB}{62,110,194}
\definecolor{xteLine}{RGB}{0,176,80}
\definecolor{dispGreen}{RGB}{0,255,46}
\definecolor{dispOrange}{RGB}{255,192,0}
\definecolor{pointText}{RGB}{237,125,49}
\useasboundingbox (-4,-4) rectangle (1336,524);   

\newcommand{\plate}[4]{%
  \fill[plateFill] (#1) -- (#4) -- ($(#4)+(0,10.5)$) -- ($(#1)+(0,10.5)$) -- cycle;
  \fill[plateFill] (#4) -- (#3) -- ($(#3)+(0,10.5)$) -- ($(#4)+(0,10.5)$) -- cycle;
  \fill[plateFill] (#1) -- (#2) -- (#3) -- (#4) -- cycle;
  \foreach \f in {0.346,0.691} \draw[meshGrey,line width=1.8pt] ($(#1)!\f!(#2)$) -- ($(#4)!\f!(#3)$);   
  \foreach \f in {0.303,0.655} \draw[meshGrey,line width=1.8pt] ($(#1)!\f!(#4)$) -- ($(#2)!\f!(#3)$);   
  }

\coordinate (L1) at (230.5,250.2); \coordinate (T1) at (672.5,0.9); \coordinate (R1) at (1114.5,250.2); \coordinate (B1) at (672.5,499.5);
\plate{L1}{T1}{R1}{B1}

\begin{scope}[plateEdge,line width=1.5pt,line cap=round]
\draw (230.5,250.2) -- (672.5,0.9);
\draw (672.5,0.9) -- (1114.5,250.2);
\draw (1114.5,250.2) -- (672.5,499.5);
\draw (672.5,499.5) -- (230.5,250.2);
\draw (230.5,260.7) -- (672.5,510.0);
\draw (672.5,510.0) -- (1114.5,260.7);
\draw (230.5,250.2) -- (230.5,260.7);
\draw (672.5,499.5) -- (672.5,510.0);
\draw (1114.5,250.2) -- (1114.5,260.7);
\end{scope}

\fill[red,opacity=0.4]       (219.4,245.4) -- (677.3,-13.6) -- (689.9,-4.9) -- (689.9,12.5) -- (232.0,271.5) -- (219.4,262.8) -- cycle;
\fill[dispGreen,opacity=0.4] (890.4,188.4) -- (1055.0,282.1) -- (1067.5,290.8) -- (1067.5,308.2) -- (902.9,214.4) -- (890.4,205.8) -- cycle;

\foreach \x/\ya/\yb in {
    232.5/251.4/375.0,
    295.4/215.4/326.8,
    358.2/179.5/278.6,
    421.1/143.5/230.4,
    483.9/107.5/182.1,
    546.8/71.6/133.9,
    609.6/35.6/85.7,
    672.5/0.0/37.5
}
  \draw[red,line width=5.5pt,-{Triangle[length=17pt,width=17pt]}]
      (\x,\ya) -- (\x,\yb);
\foreach \x/\ya/\yb in {910.7/206.7/95.6, 946.5/227.0/115.9, 983.3/241.5/130.4, 1019.1/264.7/153.6, 1055.9/288.9/178.7}
  \draw[dispGreen,line width=5.5pt,-{Triangle[length=17pt,width=17pt]}] (\x,\ya) -- (\x,\yb);

\fill[dispOrange] (661.8,243.0) -- (671.5,248.3) -- (671.5,263.5) -- (661.8,270.0) -- (652.2,263.5) -- (652.2,248.3) -- cycle;
\draw[dispOrange,line width=6pt,-{Triangle[length=19pt,width=20pt]}] (661.8,257.0) -- (583.4,296.6);

\draw[axisBlue,line width=4.6pt,-{Triangle[length=17pt,width=18pt]}] (672.5,510.0) -- (672.5,424.7); 
\draw[axisBlue,line width=4.6pt,-{Triangle[length=17pt,width=18pt]}] (672.5,510.0) -- (740.7,470.4);  
\draw[axisBlue,line width=4.6pt,-{Triangle[length=17pt,width=18pt]}] (672.5,510.0) -- (604.7,470.9);  

\node[lab,text=red] at (289.1,47.0) {\fs{35.8}Prescribed varying};
\node[lab,text=red] at (288.6,91.3) {\fs{35.8}Line Displ (YCTE)};
\node[lab,text=xteLine] at (1106.3,70.7) {\fs{35.8}Prescribed Line};
\node[lab,text=xteLine] at (1105.3,115.0) {\fs{35.8}Displ (XCTE)};
\node[lab,text=pointText] at (588.3,330.8) {\fs{35.8}Point Displ};
\node[lab,text=axisBlue] at (637.7,441.5) {\fs{34}$\bm{z}$};
\node[lab,text=axisBlue] at (765.7,493.5) {\fs{34}$\bm{x}$};
\node[lab,text=axisBlue] at (590.7,506.2) {\fs{34}$\bm{y}$};
\end{tikzpicture}
}
\caption{}
\end{subfigure}
\caption{(a) Connections of the multidomain assembly: along an edge at constant $x$ (XCTE), $y$ (YCTE) and over a common area (AR). (b) Prescribed displacements at a point and along lines at constant $x$ or $y$.}
\label{fig:connections}
\end{figure}

The domains are connected by penalties~\citep{Saullo_MD_Delam}, Fig.~\ref{fig:connections}a. Two domains $i$ and $j$ that share an edge at constant $x$ (\conn{SSxcte}) are connected by the energy
\begin{equation}\label{eq:pen-edge}
\begin{aligned}
U^{ij}_{e} = \frac{1}{2}\int_0^{b} \Big[ &k_t \big( (u^i - u^j)^2 + (v^i - v^j)^2 \\[-2pt]
&+ (w^i - w^j)^2 \big) + k_r \left( w^i_{,x} - w^j_{,x} \right)^2 \Big] \dd y,
\end{aligned}
\end{equation}
evaluated along the common edge. The default penalties of Castro and Donadon~\citep{Saullo_MD_Delam} are derived from the laminates, $k_t = 4A^i_{11}A^j_{11}/[(A^i_{11} + A^j_{11})(h_i + h_j)]$ and $k_r = 4D^i_{11}D^j_{11}/[(D^i_{11} + D^j_{11})(h_i + h_j)]$, i.e.\ $A_{11}/h$ and $D_{11}/h$ for equal laminates. They were proposed for buckling and vibration analyses. Per unit width, the connection is a rotational spring of stiffness $k_r$, which adds to a cantilever of length $a$ loaded at its tip a compliance equal to $3D_{11}/(k_r a) = 3h/a$ of its bending compliance, about 13\% for the arm of the reference DCB ($h = 2.1$~mm, $a = 48$~mm). The connections of the arms sit at the crack front, where the moment is largest, and the default penalties lower the elastic stiffness of the DCB model of the thesis by 20\% (Appendix~\ref{app:edge}); the factor 100 brings it within 0.3\% of the converged value, and this factor is applied to the penalty constants throughout the present study. A perfect bond between two stacked domains (\conn{SB}) penalises the relative displacement of the contacting surfaces over the connection interface \citep{Saullo_MD_Delam}.

Displacements are prescribed by penalties, Fig.~\ref{fig:connections}b. For an out-of-plane displacement $w_p(y)$ imposed along the line $x = x_p$ of a domain,
\begin{equation}\label{eq:pen-pd}
\begin{aligned}
U_{PD} &= \frac{1}{2}\int_0^{b} k_w \left( w(x_p, y) - w_p(y) \right)^2 \dd y, \\
\vect{f}_{PD} &= \int_0^{b} k_w\, w_p\, \vect{S}_w^\top \dd y,\\
\mat{K}^{pen}_{PD} &= \int_0^{b} k_w \vect{S}_w^\top \vect{S}_w \dd y, 
\end{aligned}
\end{equation}
where $\vect{S}_w$ is the row of $\mat{S}$ that corresponds to $w$, evaluated at $x = x_p$. The penalty is $k_w = 10^6$~N/mm$^2$ in all analyses. The load that a load cell measures is the reaction of the prescribed displacement,
\begin{equation}\label{eq:reaction}
P = \int_0^{b} k_w \left( w_p(y) - w(x_p, y) \right) \dd y,
\end{equation}
which is used for all the load--displacement curves.

\subsubsection{Cohesive interface}\label{sec:cohesive}

A cohesive interface (\conn{SB\_TSL}) connects a top domain $t$ and a bottom domain $b$ with the same planform, whose contacting faces are at the distances $d^t$ and $d^b$ from their mid-planes; for the DCB, $d^t$ and $d^b$ are half the thicknesses of the arms. In CLPT the in-plane displacement at the face of a plate is $u \mp d\, w_{,x}$, so the displacement jump across the interface is, Fig.~\ref{fig:dcb-model}b,
\begin{equation}\label{eq:jump}
\begin{aligned}
\Delta_u &= u^t + d^t w^t_{,x} - u^b + d^b w^b_{,x}, \\
\Delta_v &= v^t + d^t w^t_{,y} - v^b + d^b w^b_{,y}, \\
\Delta_w &= w^t - w^b,
\end{aligned}
\end{equation}
with the slope of each domain. Inside the fracture process zone the two arms of a DCB rotate in opposite directions, and the slope of one domain cannot represent both. With Eq.~\eqref{eq:ritz}, the separations are linear in the Ritz constants, $\Delta_\alpha = \vect{B}_\alpha(x, y)\vect{c}$ for $\alpha = u, v, w$, where the row $\vect{B}_\alpha$ is built from the approximation functions of both domains and their derivatives, e.g.\ $w_{,x} = \sum_{i,j} (2/a_p) f^w_{i,\xi}\, g^w_j\, c^w_{ij}$, resulting in

\begin{equation}
\begin{aligned}
B_{i,j} &=
\begin{bmatrix}
\cdots B_{u_{i,j}} \cdots \\
\cdots B_{v_{i,j}} \cdots \\
\cdots B_{w_{i,j}} \cdots
\end{bmatrix} \\ &=
\begin{bmatrix}
f^{u,t}_i g^{u,t}_j - f^{u,b}_i g^{u,b}_j
& 0 &
\frac{2}{a_p}[
\begin{aligned}[t]
&d^t f^{w,t}_{i,\xi} g^{w,t}_j \\
&+ d^b f^{w,b}_{i,\xi} g^{w,b}_j]
\end{aligned}
\\[18pt]
f^{v,t}_i g^{v,t}_j - f^{v,b}_i g^{v,b}_j
& 0 &
\frac{2}{b_p}[
\begin{aligned}[t]
&d^t f^{w,t}_i g^{w,t}_{j,\eta} \\
&+ d^b f^{w,b}_i g^{w,b}_{j,\eta}]
\end{aligned}
\\[18pt]
0 & 0 &
f^{w,t}_i g^{w,t}_j - f^{w,b}_i g^{w,b}_j
\end{bmatrix}
\end{aligned}
\end{equation}

\begin{figure}
\centering
\includegraphics[width=\columnwidth]{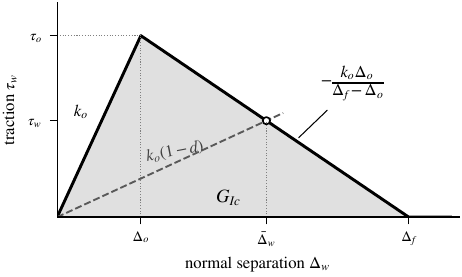}
\caption{Bilinear traction--separation law. The secant stiffness $k_o(1-d)$ gives the traction of a damaged point and its unloading path; the slope of the softening branch, $-k_o\Delta_o/(\Delta_f - \Delta_o)$, is the normal stiffness of the consistent tangent at a loading point, Eq.~\eqref{eq:tangent}. The shaded area is $G_{Ic}$.}
\label{fig:tsl}
\end{figure}

\paragraph{Traction--separation law} The tractions follow a bilinear law with one damage variable $d$ that degrades the three components, Fig.~\ref{fig:tsl}:
\begin{equation}\label{eq:law}
\tau_\alpha = k_o \left(1 - d\right) \Delta_\alpha, \qquad \alpha = u, v, w,
\end{equation}
where $k_o$ is the initial stiffness. The onset and failure separations are $\Delta_o = \tau_o/k_o$ and $\Delta_f = 2G_{Ic}/\tau_o$, with $\tau_o$ the interface strength and $G_{Ic}$ the mode~I fracture toughness, and the damage is driven by the normal separation and is irreversible:
\begin{equation}\label{eq:damage}
\begin{gathered}
d = \max\left( d^{h}, \hat d(\bar\Delta_w) \right), \\[4pt]
\hat d(\Delta) = \begin{cases}
0 & \Delta \le \Delta_o, \\[2pt]
\dfrac{\Delta_f (\Delta - \Delta_o)}{(\Delta_f - \Delta_o) \Delta} & \Delta_o < \Delta < \Delta_f, \\[8pt]
1 & \Delta \ge \Delta_f,
\end{cases}
\end{gathered}
\end{equation}
where $d^h$ is the damage of the last converged state at the same point and $\bar\Delta_w$ is the separation that drives the damage, defined below. The damage history is updated only at converged states.

\paragraph{Internal force of the interface} The work of the tractions on the separations gives the internal force of the interface,
\begin{equation}\label{eq:fcoh}
\begin{aligned}
\vect{f}^{coh}_{int} &= \int_A k_o (1 - d) \sum_{\alpha} \vect{B}_\alpha^\top \Delta_\alpha \dd A = \mat{K}^{coh}_s(d)\vect{c}, \\
\mat{K}^{coh}_s(d) &= \int_A k_o (1 - d) \sum_{\alpha} \vect{B}_\alpha^\top \vect{B}_\alpha \dd A,
\end{aligned}
\end{equation}
where $\mat{K}^{coh}_s$ is the secant stiffness of the interface; $\alpha$ runs over $u, v, w$ in all sums. The integrals are evaluated with a Gauss--Legendre rule of $n_x \times n_y$ points, at which the damage is stored. As the traction of each point follows the law of Eq.~\eqref{eq:law}, the work dissipated by a point that fails under an increasing opening is the area under the law, $G_{Ic}$: the dissipation is already contained in the degradation of the stiffness, and no energy of crack formation is added to the total potential (Appendix~\ref{app:ucrack}).

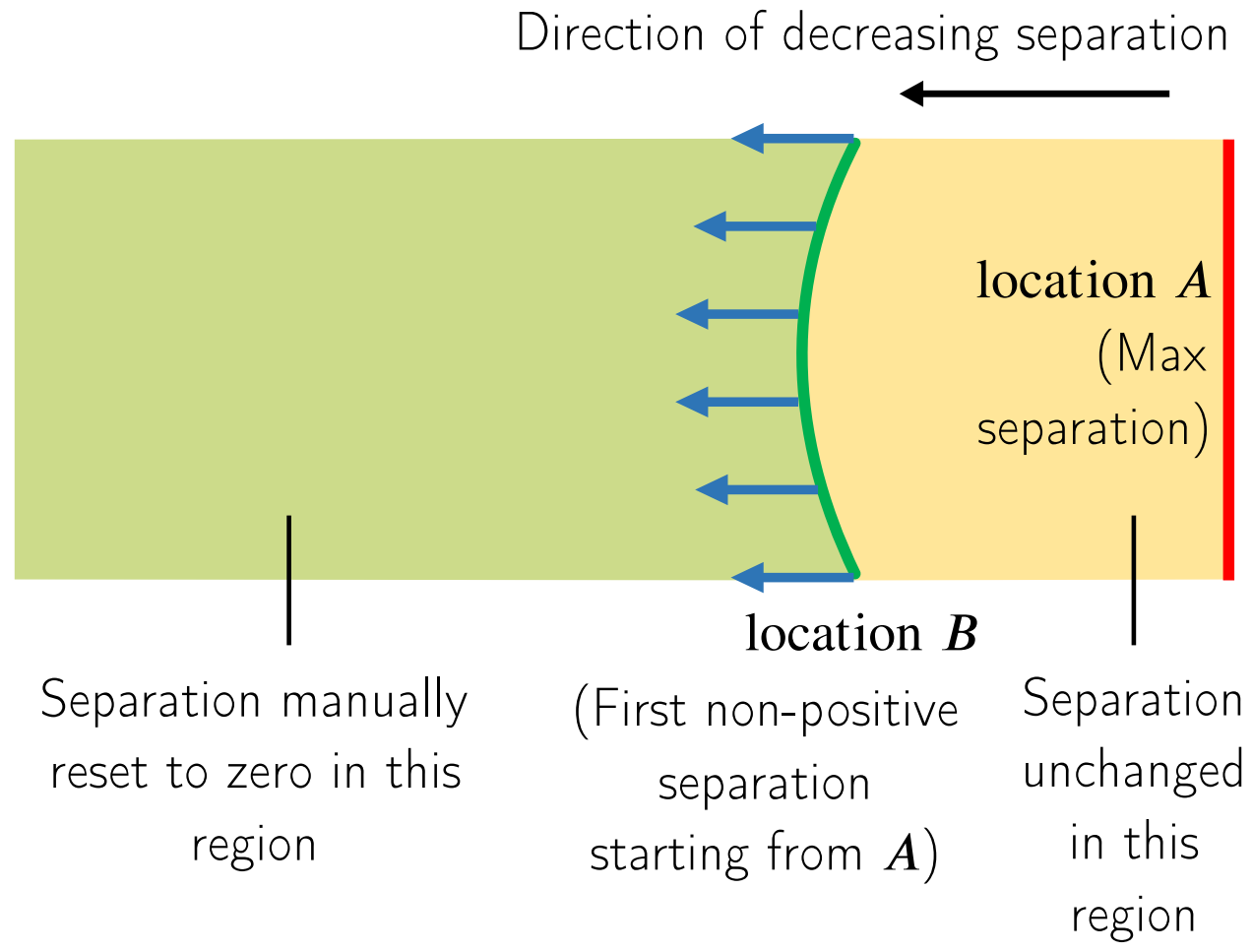
\begin{figure}
\centering
 \resizebox{\columnwidth}{!}{%
\begin{tikzpicture}[x=1pt,y=-1pt,                 
  line cap=butt, line join=round,
  lab/.style={anchor=base,inner sep=0pt,outer sep=0pt,text height=0pt,text depth=0pt}]

\definecolor{regGreen}{RGB}{204,219,138}   
\definecolor{regYellow}{RGB}{255,230,153}  
\definecolor{curveGreen}{RGB}{0,176,80}
\definecolor{arrowBlue}{RGB}{46,117,182}
\useasboundingbox (-4,-4) rectangle (659,513);   

\fill[regGreen]  (5.5,76) -- (455.3,75.5) .. controls (413.6,153.8) and (416.6,232.2) .. (452.7,310.5) -- (5.5,310) -- cycle;
\fill[regYellow] (651.5,76) -- (455.3,75.5) .. controls (413.6,153.8) and (416.6,232.2) .. (452.7,310.5) -- (651.5,310) -- cycle;

\draw[curveGreen,line width=6pt,line cap=round] (452.7,78) .. controls (413.6,153.8) and (416.6,232.2) .. (452.7,307);
\draw[red,line width=6pt] (651.5,76) -- (651.5,310.5);

\foreach \y/\xs in {
    75.5/452,
    122.2/432.0,
    168.9/422.5,
    215.6/422.5,
    262.3/433.0,
    309/452.0
}
  \draw[arrowBlue,line width=5pt,-{Triangle[length=18pt,width=15pt]}]
      (\xs,\y) -- ({\xs-66},\y);

\draw[black,line width=3.6pt,-{Triangle[length=13pt,width=13pt]}] (620,51.7) -- (476,51.7);

\draw[black,line width=2.4pt] (151.5,276) -- (151.5,345);   
\draw[black,line width=2.4pt] (601,276) -- (601,345);       

\node[lab,text=black] at (462.0,28.3) {\fs{31.7}Direction of decreasing separation};
\node[lab,text=black] at (580.0,160.0) {\fs{30.8}$\mathrm{location}\ A$};   
\node[lab,text=black] at (610.0,199.0) {\fs{29.8}(Max};
\node[lab,text=black] at (580.0,238.0) {\fs{31.3}separation)};
\node[lab,text=black] at (405.0,387.8) {\fs{31.4}(First non-positive};
\node[lab,text=black] at (405.0,426.8) {\fs{31.1}separation};
\node[lab,text=black] at (405.0,464.4) {\fs{30.8}starting from $A$)};
\node[lab,text=black] at (456.5,347.6) {\fs{30.4}$\mathrm{location}\ B$};
\node[lab,text=black] at (133.0,382.9) {\fs{31.5}Separation manually};
\node[lab,text=black] at (133.5,420.8) {\fs{31.4}reset to zero in this};
\node[lab,text=black] at (133.0,459.3) {\fs{30.1}region};
\node[lab,text=black] at (601.0,382.3) {\fs{31.3}Separation};
\node[lab,text=black] at (601.0,420.2) {\fs{31.4}unchanged};
\node[lab,text=black] at (601.0,458.6) {\fs{31.2}in this};
\node[lab,text=black] at (601.0,497.0) {\fs{31.2}region};
\end{tikzpicture}%
}
\caption{Correction of the separation that drives the damage, for a single crack front: from the maximum separation, $A$, the separation is kept up to the first non-positive value, $B$, and set to zero beyond it.}
\label{fig:correction}
\end{figure}

\paragraph{Separation that drives the damage} The opening ahead of the process zone of a DCB decays with the oscillations of two beams connected by an elastic foundation, with the rate $\lambda$ of Eq.~\eqref{eq:lcz}: its first negative lobe is physical, and the global Ritz functions add small positive lobes further ahead, which grow with the load. When used directly to compute the damage, they create isolated bands of damage far from the crack front. The damage is therefore computed from a corrected separation, for a single crack front: for each row $i$ of integration points at constant $y$, starting from the point of maximum separation, $A_i$, and moving along the direction of decreasing separation, the first non-positive separation is found, $B_i$. The separation is set to zero from $B_i$ to the edge of the domain, Fig.~\ref{fig:correction}. For a crack front that advances along $-x$, with $j$ the column of the grid of points, increasing with $x$,
\begin{equation}\label{eq:correction}
\begin{gathered}
\bar\Delta_{w,ij} = \begin{cases}
0 & j \le B_i, \\
\Delta_{w,ij} & j > B_i,
\end{cases} \\[4pt]
B_i = \max \left\{ j \le A_i : \Delta_{w,ij} \le 0 \right\}, \\[2pt]
A_i = \arg\max_j \Delta_{w,ij},
\end{gathered}
\end{equation}
If a row has no non-positive separation for $j\leq A_i$, $B_i$ is not defined and the row is not modified. The corrected separation is used only to compute the damage; the tractions of Eq.~\eqref{eq:fcoh} use the full separation, including its compressive part, which balances the tensile tractions of the process zone.

\paragraph{Residual} The residual of the assembly is
\begin{equation}\label{eq:residual}
\vect{R}(\vect{c}) = \vect{f}^{pan}_{int}(\vect{c}) + \mat{K}^{pen}_{conn}\vect{c} + \mat{K}^{coh}_s(d(\vect{c}))\vect{c} + \mat{K}^{pen}_{PD}\vect{c} - \vect{f}_{PD},
\end{equation}
where $\mat{K}^{pen}_{conn}$ gathers the edge and perfect-bond connections, and $d(\vect{c})$ is evaluated with Eqs.~\eqref{eq:damage} and~\eqref{eq:correction} at every evaluation of the residual: the secant stiffness of the interface is part of the internal force and cannot be frozen between iterations (Appendix~\ref{app:secant}). Equilibrium is reached when the ratio of scaled norms~\citep{Crisfield1981}
\begin{equation}\label{eq:criterion}
\begin{gathered}
\frac{F(\vect{R})}{\max\left( F(\vect{f}_{int}), F(\mat{K}^{pen}_{PD}\vect{c} - \vect{f}_{PD}) \right)} < \varepsilon, \\[4pt]
F(\vect{x}) = \sqrt{\vect{x}^\top \mat{D}^{-1} \vect{x}},
\end{gathered}
\end{equation}
is below $\varepsilon = 10^{-4}$, where $\vect{f}_{int} = \vect{f}^{pan}_{int} + \mat{K}^{pen}_{conn}\vect{c} + \vect{f}^{coh}_{int}$ is the internal force, $\mat{K}^{pen}_{PD}\vect{c} - \vect{f}_{PD}$ is the reaction of the prescribed displacement, and $\mat{D}$ holds the absolute values of the diagonal of the iteration matrix at the start of the increment. The denominator measures the physical forces, of the order of $10^2$~N, and not the penalty terms $\mat{K}^{pen}_{PD}\vect{c}$ and $\vect{f}_{PD}$ separately, which are of the order of $10^6$--$10^7$ for $k_w = 10^6$~N/mm$^2$ (Appendix~\ref{app:criterion}).

\subsubsection{Consistent tangent of the damaged interface}\label{sec:tangent}

The derivative of the internal force of Eq.~\eqref{eq:fcoh} with respect to the Ritz constants contains, besides the secant stiffness, the change of the damage with the separation:
\begin{equation}\label{eq:tangent}
\begin{aligned}
\frac{\partial \vect{f}^{coh}_{int}}{\partial \vect{c}} &= \mat{K}^{coh}_s + \mat{K}^{coh}_{\dot d}, \\
\mat{K}^{coh}_{\dot d} &= - \int_A \chi\, k_o\, \hat d'(\Delta_w) \left( \sum_\alpha \vect{B}_\alpha^\top \Delta_\alpha \right) \vect{B}_w \dd A,
\end{aligned}
\end{equation}
with 
\begin{equation}\label{eq:dprime}
\begin{gathered}
\hat d'(\Delta) = \frac{\Delta_f \Delta_o}{(\Delta_f - \Delta_o) \Delta^2}, \\[4pt]
\chi = \begin{cases}
1 & \Delta_o < \bar\Delta_w < \Delta_f \ \text{and}\ \hat d(\bar\Delta_w) > d^h, \\
0 & \text{otherwise}.
\end{cases}
\end{gathered}
\end{equation}
In Eqs.~\eqref{eq:tangent} and~\eqref{eq:dprime}, $\chi$ is a mask that selects the points where the damage grows in the current state. There, the correction of Eq.~\eqref{eq:correction} leaves the separation unchanged, $\bar\Delta_w = \Delta_w$, with the derivative $\vect{B}_w$; the correction switches between 0 and $\Delta_w$, and its derivative is zero almost everywhere. The matrix $\mat{K}^{coh}_{\dot d}$ is not symmetric, because the tangential tractions depend on $d$, which in turn depends on $\Delta_w$, but is independent of $\Delta_u$ and $\Delta_v$. Its normal part, $-\chi k_o \hat d' \Delta_w \vect{B}_w^\top\vect{B}_w$, is symmetric and negative semi-definite, and at a loading point the normal stiffness of the tangent becomes $k_o(1 - d - \hat d' \Delta_w) = -k_o\Delta_o/(\Delta_f - \Delta_o)$, the slope of the softening branch, Fig.~\ref{fig:tsl}. Using only the secant stiffness, the iteration matrix overestimates the stiffness of every softening point, and the iterations converge at best linearly once the crack grows, which on the reference DCB stops the analysis at the first opening after the peak (Section~\ref{sec:res-verification}).

\subsubsection{Assembly by matrix products}\label{sec:assembly}

Since the secant stiffness is part of the internal force, it is needed at every evaluation of the residual, including the trials of the line search of Section~\ref{sec:solver}. Assembled with loops over the terms of both domains and over the integration points, its cost grows with the product of the numbers of terms and of integration points at every evaluation. Here, the same Gauss--Legendre rule is written as matrix products. Let $g = 1, \dots, n_g$ be the integration points, with the weights $w_g = w_\xi w_\eta\, a_p b_p/4$, and $\vect{c}_{tb}$ the Ritz constants of the top domain followed by those of the bottom domain. The separations at all points are given by the $n_g \times n_{dof}$ operators
\begin{equation}\label{eq:operators}
\begin{aligned}
\mat{B}_u &= \begin{bmatrix} \mat{S}^t_u + d^t \mat{S}^t_{w,x} & -\mat{S}^b_u + d^b \mat{S}^b_{w,x} \end{bmatrix}, \\
\mat{B}_v &= \begin{bmatrix} \mat{S}^t_v + d^t \mat{S}^t_{w,y} & -\mat{S}^b_v + d^b \mat{S}^b_{w,y} \end{bmatrix}, \\
\mat{B}_w &= \begin{bmatrix} \mat{S}^t_w & -\mat{S}^b_w \end{bmatrix},
\end{aligned}
\end{equation}
where row $g$ of $\mat{S}^t_u$ holds the approximation functions of $u$ of the top domain at point $g$, and so on.
The operators depend only on the approximation functions, the integration points, the dimensions $a_p$, $b_p$ and the offsets $d^t$, $d^b$; they do not change during the analysis and are computed once and cached.
With $\mathrm{diag}(\cdot)$ the diagonal matrix of a vector over the points and $k_g = k_o(1 - d_g)$,
\begin{equation}\label{eq:K-products}
\mat{K}^{coh}_s = \sum_{\alpha} \mat{B}_\alpha^\top \mathrm{diag}\left( w_g k_g \right) \mat{B}_\alpha,
\end{equation}
a product evaluated by optimised BLAS routines. Splitting $k_g = k_o - (k_o - k_g)$,
\begin{equation}\label{eq:K-update}
\begin{aligned}
\mat{K}^{coh}_s &= \mat{K}^{coh}_o - \sum_{\alpha} {\mat{B}_\alpha}^\top_{\mathcal{D}}\, \mathrm{diag}\left( w_g (k_o - k_g) \right)_{\mathcal{D}} {\mat{B}_\alpha}_{\mathcal{D}}, \\
\mat{K}^{coh}_o &= k_o \sum_{\alpha} \mat{B}_\alpha^\top \mathrm{diag}\left( w_g \right) \mat{B}_\alpha,
\end{aligned}
\end{equation}
where $\mathcal{D} = \{ g : k_g \neq k_o \}$ is the set of damaged points and the subscript $\mathcal{D}$ keeps only their rows. The pristine stiffness $\mat{K}^{coh}_o$ is computed once for each $k_o$ and cached. Before damage onset $\mathcal{D}$ is empty and the assembly reduces to a copy of $\mat{K}^{coh}_o$; afterwards, $\mathcal{D}$ holds the process zone and the cracked region, and the cost of the update grows with the number of the damaged integration points, not with the size of the interface. Equation~\eqref{eq:K-update} is Eq.~\eqref{eq:K-products} regrouped: it is exact, and it does not depend on how the damage is computed. The tangent of Eq.~\eqref{eq:tangent} uses the same operators, restricted to the rows where $\chi = 1$. The kernels that loop over the terms and the integration points, with the kinematics of Eq.~\eqref{eq:jump}, are kept for domains with different planforms, where the functions of the bottom domain are evaluated at the natural coordinates of the top domain.

\subsubsection{Solution procedure}\label{sec:solver}

The analyses are displacement controlled, with a list of prescribed displacements $w_{p,n}$, and the equilibrium of each increment is found by a modified Newton--Raphson method~\citep{deBorst2012} with the iteration matrix
\begin{equation}\label{eq:iteration-matrix}
\mat{K} = \mat{K}_C + \mat{K}_G + \mat{K}^{pen}_{conn} + \mat{K}^{coh}_s + \mat{K}^{coh}_{\dot d} + \mat{K}^{pen}_{PD},
\end{equation}
which is not symmetric; the linear systems are solved with a sparse LU factorisation. The tangent of the domains, $\mat{K}_C + \mat{K}_G$, is refreshed every $n_{NR} = 3$ iterations, while $\mat{K}^{coh}_s$ and $\mat{K}^{coh}_{\dot d}$ are evaluated at every iteration. 

\begin{figure}
\centering
\includegraphics[width=\columnwidth]{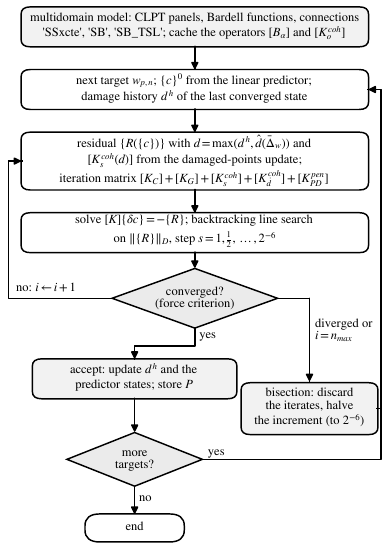}
\caption{Solution procedure for one prescribed displacement of the list: predictor, Eq.~\eqref{eq:predictor}; residual, Eqs.~\eqref{eq:residual} and~\eqref{eq:K-update}; iteration matrix, Eq.~\eqref{eq:iteration-matrix}; line search, Eq.~\eqref{eq:armijo}; convergence criterion, Eq.~\eqref{eq:criterion}.}
\label{fig:flowchart}
\end{figure}

Figure~\ref{fig:flowchart} illustrates the solution procedure. After the peak load, a pure Newton--Raphson method can stall or cycle when the integration points are too sparse. The residual of Eq.~\eqref{eq:residual} is continuous but only piecewise smooth, such that its derivative jumps whenever: an integration point crosses $\Delta_o$ or $\Delta_f$; switches between loading ($\chi = 1$) and unloading ($\hat d \le d^h$); or enters or leaves the region zeroed by Eq.~\eqref{eq:correction}. The tangent of Eq.~\eqref{eq:tangent} is exact only on one side of each kink. Before the peak, a few points cross a kink in one iteration, but after the peak, the global Ritz functions move the whole crack front at once, and a full correction can carry the front across several integration points, after which the iterations may stall or cycle between states on both sides of the kinks. The line search and the bisection below are safeguards against this; verification V4 (Section~\ref{sec:res-verification}) shows when they are needed.

\paragraph{Linear predictor} Under displacement control, a displacement increment that starts from the last converged state has a first residual dominated by the jump of the prescribed displacement. The increment $n$ therefore starts from the linear extrapolation of the last two converged states,
\begin{equation}\label{eq:predictor}
\vect{c}^{0}_{n} = \vect{c}_{n-1} + \frac{w_{p,n} - w_{p,n-1}}{w_{p,n-1} - w_{p,n-2}} \left( \vect{c}_{n-1} - \vect{c}_{n-2} \right),
\end{equation}
with the state $w_p = 0$, $\vect{c} = \vect{0}$ used for the second increment. The damage of the predicted state is not stored. While no point is damaged, $\mat{K}_C + \mat{K}_G$ is reused for up to five increments, and within an increment it is refreshed every $n_{NR}$ iterations but not after the first one; after damage onset it is evaluated at the start of every increment.

\paragraph{Backtracking line search} With $\vect{\delta c}$ the solution of $\mat{K}\vect{\delta c} = -\vect{R}(\vect{c}_i)$, the new iterate is $\vect{c}_{i+1} = \vect{c}_i + s\vect{\delta c}$, with the step $s$ the first of $1, \frac{1}{2}, \frac{1}{4}, \dots, 2^{-6}$ that satisfies
\begin{equation}\label{eq:armijo}
\begin{gathered}
\left\| \vect{R}(\vect{c}_i + s\vect{\delta c}) \right\|_D \le \left(1 - 10^{-4} s\right) \left\| \vect{R}(\vect{c}_i) \right\|_D, \\[2pt]
\left\| \vect{x} \right\|_D = F(\vect{x}),
\end{gathered}
\end{equation}
with the scaling of Eq.~\eqref{eq:criterion}, which balances the rows of the prescribed-displacement penalty against the others. Unlike the line searches on the energy commonly used with Newton--Raphson methods~\citep{deBorst2012}, here the residual norm is evaluated, because the damaged interface has no potential and $\mat{K}$ is not symmetric. When $\mat{K}$ is the exact Jacobian $\mat{J}$ of the residual, then $\vect{\delta c}$ is a descent direction for any positive weighting,
\begin{equation}\label{eq:descent}
\begin{aligned}
&\left. \frac{\mathrm{d}}{\mathrm{d}s} \frac{1}{2} \left\| \vect{R}(\vect{c}_i + s\vect{\delta c}) \right\|_D^2 \right|_{s=0} \\[2pt]
&\qquad = \vect{R}^\top \mat{D}^{-1} \mat{J} \vect{\delta c} = - \vect{R}^\top \mat{D}^{-1} \vect{R} < 0,
\end{aligned}
\end{equation}
so that a small enough $s$ always satisfies Eq.~\eqref{eq:armijo}. Between refreshes of $\mat{K}_C + \mat{K}_G$ this is not guaranteed, and if no trial satisfies Eq.~\eqref{eq:armijo} the one with the smallest norm is taken. Halving $s$ halves the jump of the separations and thus the number of kinks crossed per iteration. Each trial costs one residual, cheap with the assembly of Section~\ref{sec:assembly}.

\paragraph{Bisection of increments} If an increment does not converge within $n_{max} = 50$ iterations, or diverges, its iterates are discarded, and the first half of the increment is solved from the last converged state. This is applied recursively with a stack of pending targets, down to $2^{-6}$ of the original increment, after which the analysis stops. Discarding the iterations keeps consistency because the damage history is updated only at converged states. Every converged sub-increment updates the history and the states of the predictor, so that a bisected increment follows a finer loading path, as needed for an irreversible law; the results are stored only at the predefined prescribed displacements of the list $w_{p,n}$.

\subsubsection{Model of the double cantilever beam}\label{sec:dcb-models}

\begin{figure*}[!tbp]
\centering
\includegraphics[width=\textwidth]{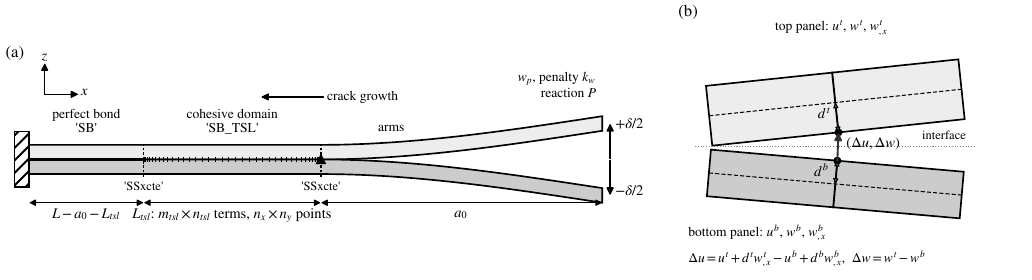}
\caption{(a) Symmetric multidomain model of the DCB used for the verification, the convergence study and the validation (not to scale): perfect bond, cohesive domain with its grid of Gauss--Legendre points, and free arms, connected along their edges; the far end is clamped and the tips are opened to $\pm\delta/2$. (b) Kinematics of the cohesive interface, Eq.~\eqref{eq:jump}: the displacement jump between the faces of the top and bottom domains uses the slope of each domain.}
\label{fig:dcb-model}
\end{figure*}

One symmetric multidomain model of the DCB is used for the verification, the convergence study and the validation, Fig.~\ref{fig:dcb-model}a. Along $x$, measured from the far end, each arm has a domain bonded by a perfect bond (\conn{SB}), which the crack does not reach, the cohesive domain of length $L_{tsl}$ (\conn{SB\_TSL}), whose edge at $x = L - a_0$ is the initial crack front, and the free arm of length $a_0$. The far end is clamped, and the tips of the arms are opened to $\pm\delta/2$ by prescribed displacements, with the rotations and the in-plane displacements free. The load is the reaction of Eq.~\eqref{eq:reaction} at the tip of the top arm. Only the region that is anticipated to be reached by the crack needs the resolution of the cohesive zone. A weld narrower than the specimen is represented by starting the integration points outside the bonded strip with $d^h = 1$, which removes the connection there while the arms keep their full width.

\paragraph{Reference DCB} The verification and the convergence study use the UdG specimen of Tijs et al.~\citep{Bas_charaterization} with $a_0 = 48$~mm, with the length $L = 100$~mm of their FE model: width $b = 25$~mm, arms $[0]_{15}$ of AS4D/PEKK-FC with plies of 0.14~mm ($h = 2.1$~mm), $E_{11} = 133.15$~GPa, $E_{22} = 10.95$~GPa (averages of the tensile and compressive moduli), $\nu_{12} = 0.316$, $G_{12} = 5.19$~GPa, $G_{Ic} = 1.12$~N/mm (the propagation value, including fibre bridging) and $\tau_o = 87$~MPa. Two penalty stiffnesses are used: case~A, $k_o = 10^5$~N/mm$^3$, the value of the validation runs (Section~\ref{sec:val-strategy}), and case~B, $k_o = 2\times10^5$~N/mm$^3$, the mode~I penalty stiffness used by Tijs et al.~\citep{Tijs2022b} for conduction-welded joints of the same material. The cohesive domain is $L_{tsl} = 24$~mm of the 52~mm ligament, and the opening reaches 8~mm in 38 increments: 23 equal increments up to $\delta = 5$~mm and increments of 0.2~mm after it.

\paragraph{Discretisation} The discretisation of the cohesive domain follows the rules obtained from the convergence study of Section~\ref{sec:res-convergence}. The Ritz basis must resolve the decay of the opening ahead of the process zone, that of the opening $w^t - w^b$ of two arms connected by an elastic foundation of stiffness $k_o$, and the integration points must sample the cohesive zone, of length~\citep{Rice1980,Turon2007}
\begin{equation}\label{eq:lcz}
\lambda = \left( \frac{k_o}{2 D_{11}} \right)^{1/4}, \qquad l_{cz} = 0.88\, \frac{E_{22} G_{Ic}}{\tau_o^2},
\end{equation}
which gives
\begin{equation}\label{eq:rules}
m_{tsl} \approx 1.25\, \lambda L_{tsl}, \qquad n_x \approx 10\, \frac{L_{tsl}}{l_{cz}},
\end{equation}
with $m_{tsl} \le 25$, $n_x \ge 4 m_{tsl}$, a heuristic determined during the development of this work for the integration of the products of the approximation functions, $n_x$ rounded up to a multiple of 10, $n_{tsl} = 10$ and $n_y = 30$, and $m = n = 8$ terms in the other domains. The factor 1.25 has a margin: the convergence study gives loads within 0.1\% of the converged ones from $m_{tsl} \approx 1.0\,\lambda L_{tsl}$, and 25 terms give the same loads as 30 to 0.02\%, so the cap of 25 terms costs no accuracy when $L_{tsl}$ is chosen accordingly. The cohesive domain covers the crack growth plus a margin of $3 l_{cz} + 8/\lambda$: the process zone and the region ahead of it where the opening decays as $e^{-\lambda x}$, the solution of beams on an elastic foundation~\citep{Kanninen1973}, to below $e^{-8}$ of its value, so that the tractions outside the cohesive domain, which verification V3 detects, are negligible (Section~\ref{sec:res-convergence}). For demonstrative purposes, the range of openings is limited such that $L_{tsl}$ stays resolved with 25 terms, i.e.\ $L_{tsl} \le 25/(1.25\lambda)$, and the last opening is the one at which the crack front of the model reaches the end of the crack growth $\Delta a$ covered by the cohesive domain, estimated from LEFM, Eq.~\eqref{eq:lefm}, with the effective crack length $a_0 + \Delta a + l_{cz} + 1/\lambda$ of the model: the process zone and the rotation that the elastic foundation allows at its root~\citep{Kanninen1973}. The corrected beam theory, whose $\chi h$ also contains the transverse shear that the CLPT arms lack, gives a shorter crack at the same opening and would let the crack front pass the cohesive domain. This allows 4--11~mm of crack growth in the cases of Section~\ref{sec:validation}.

\subsection{Verification strategy}\label{sec:verification}

The verification checks that the equations of Section~\ref{sec:formulation} are implemented correctly and that the discretisation and the solution procedure converge. V1 and V2 use two stacked domains connected by the cohesive interface, as in the tests of \panels, and V3--V5 the reference DCB of Section~\ref{sec:dcb-models}:
\begin{enumerate}[V1.]
\item \textbf{Stiffness of the interface.} The quadratic form $\vect{c}^\top\mat{K}^{coh}_s\vect{c}$ of the connection kernels is compared with the Gauss--Legendre integral of $\sum_\alpha k_g \Delta_\alpha^2$ evaluated from the displacement fields, Eq.~\eqref{eq:jump}, for random Ritz constants, random boundary flags, domains with different numbers of terms and thicknesses, both orders of the domains in the assembly, and pristine and arbitrary damage fields. The assembly by matrix products, Eqs.~\eqref{eq:K-products}--\eqref{eq:K-update}, is compared with the kernels for damage fields with pristine, softening and failed points. Agreement to round-off precision is expected.
\item \textbf{Consistent tangent.} The matrix $\mat{K}^{coh}_s + \mat{K}^{coh}_{\dot d}$ is compared with central finite differences of $\vect{f}^{coh}_{int}(\vect{c}) = \mat{K}^{coh}_s(\vect{c})\vect{c}$ in states with softening points, with and without points where the damage history is above the current damage. A relative error of the order of the finite difference error is expected, and the secant matrix alone serves as a negative control.
\item \textbf{Equilibrium of the load.} The reaction of Eq.~\eqref{eq:reaction} is compared with the integral of the normal tractions of the interface with the full separation, $\int_A k_o(1-d)\Delta_w \dd A$, which must be equal by equilibrium of the loaded arm when the cohesive domain carries the whole load. In the model of Section~\ref{sec:dcb-models} this holds when the tractions of the perfect bond beyond the cohesive domain are negligible, i.e.\ when $L_{tsl}$ contains the process zone and the compressive lobe ahead of it, so V3 also checks the length of the cohesive domain.
\item \textbf{Solution procedure.} The number of iterations per increment and the bisections are compared between the secant and the consistent tangent, with and without the line search and the bisection, on the reference discretisation and on a discretisation with too few integration points ($n_x = 60$); the runs without bisection allow up to 200 iterations per opening and stop at the first opening that does not converge. The wall time of the reference analysis is measured alone on the machine.
\item \textbf{Convergence of the discretisation.} On the reference DCB, cases~A and~B, a full grid $m_{tsl} \in \{10, 15, 20, 25, 30\} \times n_x \in \{60, 120, 180\}$ is run with $L_{tsl} = 24$~mm, and, from the reference discretisation $m_{tsl} = 25$, $n_x = 180$, $n_{tsl} = 10$, $n_y = 30$, $m = n = 8$, the terms along $y$ ($n_{tsl} = 10 \to 15$), the Gauss points along $y$ ($n_y = 30 \to 60$), the terms of the other domains ($m = n = 8 \to 12$), the increment after $\delta = 5$~mm (halved) and the length of the cohesive domain ($L_{tsl} = 24 \to 17$~mm, also with $m_{tsl} = 15$ and $n_x = 120$) are varied. The metrics are the peak load $P_{max}$ and the opening at the peak, the loads at $\delta = 6$, 7 and 8~mm, the largest difference of the load to that of the finest run ($m_{tsl} = 30$, $n_x = 180$) over all openings, relative to its peak load, $e_P$, the largest relative difference between the reaction and the traction integral of V3, $e_{V3}$, the coefficient of variation of the load drop per increment after the peak, $CV = \mathrm{std}(\Delta P)/|\mathrm{mean}(\Delta P)|$, which is small for a smooth softening and large for a staircase or sawtooth response, the number of bisections, the number of iterations and the wall time. The rules of Eq.~\eqref{eq:rules} are derived from this study.
\end{enumerate}

\subsection{Validation strategy}\label{sec:val-strategy}

\paragraph{References and cases} The validation uses references that report at least one pure mode~I DCB case, Table~\ref{tab:cases}. For each reference, the geometry, the material, the parameters of the interface, the type of the reference model and its results and the experimental data were collected (Appendix~\ref{app:repro}):
\begin{itemize}
\item Alfano and Crisfield~\citep{Alfano2001}: DCB of Table~I with the nominal strength $t_o = 57$~MPa and with the low strength $t_o = 1.7$~MPa, for which the process zone is long (about 15~mm), and the XAS-913C DCB of Table~II, compared with the experiments of Robinson and Song~\citep{RobinsonSong1992}; 2D plane-strain interface elements.
\item Turon et al.~\citep{Turon2007}: T300/977-2 DCB with cohesive elements of 0.125~mm (the converged mesh of their study) and the experimental curve of their Fig.~5.
\item Krueger~\citep{Krueger2008}: the two pure mode~I DCB benchmarks of NASA/TM-2008-215123, a unidirectional $[0]_{24}$ specimen of T300/1076 and a multidirectional D$\pm$30 specimen of C12K/R6376, both modelled with three-dimensional solid elements and the virtual crack closure technique (VCCT). The reference has no cohesive law, no interfacial strength and no experimental load--displacement curve; its benchmark is the critical load--opening curve for the onset of delamination at mid-width (their Eq.~(5) and Figs.~14 and~37).
\item Tijs et al.~\citep{Bas_charaterization}: AS4D/PEKK-FC DCB with $a_0 = 48$~mm and 40~mm. Their FE model uses the same geometry, material and bilinear law ($G_{Ic} = 1.12$~N/mm, $\tau_o = 87$~MPa) as \panels{} for $a_0 = 48$~mm, which makes it the most direct validation, and this case is also the reference DCB of the verification (Section~\ref{sec:dcb-models}); for $a_0 = 40$~mm, without an FE curve, \panels{} uses the toughness that the test reaches after its peak, 0.95~N/mm (see Section~\ref{sec:validation}); they also show a tabular multilinear law fitted to the test.
\item Leci\~{n}ana et al.~\citep{Lecinana2023}: a fatigue study of AS4D/PEKK-FC without quasi-static load--displacement curves; only the measured initial compliance of five specimens with known crack lengths (their Table~3 and Fig.~10), a linear elastic mode~I result, is used.
\item Tijs~\citep{Tijs2023PhD}, Chapter~5: conduction-welded quasi-isotropic DCB of AS4D/PEKK-FC, $[0/45/90/-45/0/-45/90/45]_s$ arms of 2.24~mm, in two configurations, ``full weld'', 25~mm wide with a weld of about 15~mm, and ``weld center'', 12.7~mm wide, cut from the middle of the weld, with FE results and tests for $G_{Ic} = 2.1$~N/mm, and the compliance calibration of his Table~5.4.1 for the autoclave unidirectional and the two welded configurations~\citep{skin_stiff_fig_bas}.
\end{itemize}

\begin{table*}[!tbp]
\caption{Mode~I DCB validation cases. $E_{11}$, $E_{22}$, $G_{12}$ in GPa; $h$ thickness of one arm, $b$ width, $a_0$ initial crack length, $L$ length, in mm; $G_{Ic}$ in N/mm; $\tau_o$ in MPa; $k_o$ of the reference in N/mm$^3$. The last three columns give the discretisation of the cohesive domain of \panels, with $k_o = 10^5$~N/mm$^3$: length $L_{tsl}$ (mm), terms $m_{tsl}$ and Gauss points $n_x$.}\label{tab:cases}
\footnotesize
\setlength{\tabcolsep}{3pt}
\resizebox{\linewidth}{!}{%
\begin{tabular}{@{}lllrrrrrrrrrlrrr@{}}
\toprule
Reference & Material & Arms & $E_{11}$ & $E_{22}$ & $G_{12}$ & $h$ & $b$ & $a_0$ & $L$ & $G_{Ic}$ & $\tau_o$ & $k_o$ (ref.) & $L_{tsl}$ & $m_{tsl}$ & $n_x$ \\
\midrule
Alfano and Crisfield~\citep{Alfano2001}, Table~I & -- & UD & 135.3 & 9.0 & 5.2 & 1.5 & 20 & 30 & 100 & 0.28 & 57 & $5.7\times10^8$ & 18.7 & 25 & 280 \\
Alfano and Crisfield~\citep{Alfano2001}, Table~I & -- & UD & 135.3 & 9.0 & 5.2 & 1.5 & 20 & 30 & 100 & 0.28 & 1.7 & $1.7\times10^7$ & 70.0$^\S$ & 30 & 120 \\
Alfano and Crisfield~\citep{Alfano2001}, Table~II & XAS-913C & UD & 126.0 & 7.5 & 4.98 & 1.5 & 30 & 30 & 100 & 0.281 & 57 & $5.7\times10^7$ & 18.4 & 25 & 330 \\
Turon et al.~\citep{Turon2007} & T300/977-2 & UD & 150.0 & 11.0 & 6.0 & 1.98 & 20 & 55 & 150 & 0.352 & 60 & $10^6$ & 23.6 & 25 & 250 \\
Krueger~\citep{Krueger2008} & T300/1076 & UD & 139.4 & 10.16 & 4.6 & 1.5 & 25 & 30 & 150 & 0.1703$^\dagger$ & 40, 60$^\ddagger$ & -- & 18.9 & 25 & 200, 450 \\
Krueger~\citep{Krueger2008} & C12K/R6376 & MD & 146.9 & 10.6 & 5.45 & 2.0 & 25.4 & 31 & 177.8 & 0.3405 & 40, 60$^\ddagger$ & -- & 21.8 & 25 & 110, 250 \\
Tijs et al.~\citep{Bas_charaterization} & AS4D/PEKK-FC & UD & 133.15 & 10.95 & 5.19 & 2.1 & 25 & 48, 40 & 100 & 1.12, 0.95$^\P$ & 87 & $10^6$ & 24.0 & 25 & 170, 200 \\
Leci\~{n}ana et al.~\citep{Lecinana2023} & AS4D/PEKK-FC & UD & 138.3 & 10.4 & 5.19 & 2.07 & 25 & 32.05--42.97 & $a_0 + 100$ & -- & -- & -- & \multicolumn{3}{c}{linear analysis} \\
Tijs~\citep{Tijs2023PhD}, full weld & AS4D/PEKK-FC & QI & 133.15 & 10.95 & 5.19 & 2.24 & 25$^\ast$ & 44.3 & 144.3 & 2.1 & 87 & $10^6$ & 21.7 & 25 & 100 \\
Tijs~\citep{Tijs2023PhD}, weld center & AS4D/PEKK-FC & QI & 133.15 & 10.95 & 5.19 & 2.24 & 12.7 & 36.2 & 136.2 & 2.1 & 87 & $10^6$ & 21.7 & 25 & 100 \\
\bottomrule
\end{tabular}}
\\[2pt]{\footnotesize $^\ast$ weld of 15~mm. $^\S$ with $k_o = 10^3$~N/mm$^3$, see Section~\ref{sec:res-convergence}. $^\P$ $a_0 = 48$ and 40~mm, the latter with the propagation toughness of the test ``DCB (2)'' (see text). $^\dagger$ toughness of T300/914C, the elastic properties being those of T300/1076, as in Section~5.1 of the reference. $^\ddagger$ the reference uses VCCT and gives neither $\tau_o$ nor $k_o$; the two values are choices (see text). UD: $[0]$; MD: one arm of the D$\pm$30 laminate, $[30/-30/0/-30/0/30/0_4/30/0/-30/0/-30/30]$ of 0.125~mm plies; QI: $[0/45/90/-45/0/-45/90/45]_s$ of 0.14~mm plies. $\nu_{12}$: 0.24, 0.263, 0.25, 0.30, 0.33 and 0.316 for Alfano and Crisfield (Table~I, Table~II), Turon et al., Krueger (T300/1076, C12K/R6376) and AS4D/PEKK-FC. The length of the welded specimens beyond the crack front (100~mm) and their elastic properties are assumed (see text).}
\end{table*}

\paragraph{Metrics and references of LEFM} The load--opening curves are compared through the initial stiffness $K_0$ (least squares through the origin of the points below 60\% of the opening at the peak), the peak load $P_{max}$ and the opening at the peak, and the load after the peak; the initial compliance is compared with the crack length for Leci\~{n}ana et al.\ and Tijs. Besides the FE and the experimental results, the reference of LEFM is beam theory under displacement control, linear up to the critical opening of $a_0$ and then following the propagation branch $G = G_{Ic}$:
\begin{equation}\label{eq:lefm}
P = \frac{\sqrt{G_{Ic}\, b_G\, E I}}{a_e}, \qquad \delta = \frac{2 P a_e^3}{3 E I}, \qquad a_e = a + \chi h,
\end{equation}
with $EI = D_{11} b$ the bending stiffness of an arm, $b_G$ the bonded width, and $\chi = 0$ for simple beam theory (SBT) or, for corrected beam theory (CBT), the correction of Williams~\citep{Williams1989} in the form given by Hashemi et al.~\citep{analytical_dcb},
\begin{equation}\label{eq:chi}
\begin{gathered}
\chi = \sqrt{\frac{E_{11}}{11 G_{13}}\left[ 3 - 2\left( \frac{\Gamma}{1 + \Gamma} \right)^2 \right]}, \\[4pt]
\Gamma = 1.18\,\frac{\sqrt{E_{11} E_{22}}}{G_{13}},
\end{gathered}
\end{equation}
with $G_{13} = G_{12}$ and, for the multidirectional and quasi-isotropic arms, the bending modulus $12 D_{11}/h^3$ in place of $E_{11}$.

\paragraph{Conventions and data} Alfano and Crisfield plot the reaction against the displacement $u$ of one arm, so $\delta = 2u$; the displacement of Turon et al.\ is taken as the opening between the loading points. Values given in tables of the references are used as such: 
the crack lengths of Leci\~{n}ana et al.\ and the compliance calibration of Tijs. The other curves were read from the figures by eye and are approximate, with an uncertainty of about 2--5~N. The test ``DCB (2)'' of Tijs et al.~\citep{Bas_charaterization} has $a_0 = 40$~mm. Where a reference is incomplete, the following is assumed: the elastic properties of the welded specimens, taken as those of Tijs et al.~\citep{Bas_charaterization}, and their length beyond the crack front, 100~mm.

\paragraph{Model settings} The model of Section~\ref{sec:dcb-models} is used with the discretisation of Eq.~\eqref{eq:rules}, Table~\ref{tab:cases}, and 60 openings, except for the low-strength case of Alfano and Crisfield, for which the cohesive domain covers the whole bonded length with 30 terms, $k_o = 10^3$~N/mm$^3$ and 40 openings up to 4~mm (Section~\ref{sec:res-convergence}). Otherwise the penalty stiffness is $k_o = 10^5$~N/mm$^3$ instead of the values of the references: Turon et al.~\citep{Turon2007} showed that the response of the DCB does not change for $k_o > 10^4$~N/mm$^3$ (their Fig.~9), and a lower $k_o$ gives a longer decay length $1/\lambda$, which the Ritz basis resolves with fewer terms. The case of Turon et al.\ is also run with their $k_o = 10^6$~N/mm$^3$ as a sensitivity check.

\section{Results and discussion}\label{sec:results}

\subsection{Verification}\label{sec:res-verification}

\paragraph{Stiffness and tangent of the interface (V1, V2)} The tests use two domains of $17 \times 25$~mm with different laminates (12 and 9 plies of 0.14~mm) and numbers of terms ($6 \times 4$ and $5 \times 5$), and $20 \times 12$ integration points. For V2 the states are scaled such that the largest normal separation is 0.015~mm, inside the softening range of the law with $k_o = 5\times10^4$~N/mm$^3$, $\tau_o = 87$~MPa and $G_{Ic} = 1.12$~N/mm ($\Delta_o = 0.0017$~mm, $\Delta_f = 0.026$~mm), such that 38--54\% of the points are softening. Table~\ref{tab:verification} gives the results. The kernels with the kinematics of Eq.~\eqref{eq:jump} reproduce the energy of the interface to round-off, for pristine and damaged interfaces, and so does the assembly by matrix products, Eq.~\eqref{eq:K-update}, which confirms that the update restricted to the damaged points is exact. The consistent tangent matches the finite differences to the accuracy of the differences, also where the damage history is above the current damage and $\chi = 0$, while the secant matrix alone is 8--62\% off.

\begin{table*}[!tbp]
\caption{Verification of the interface and of the load. Relative differences; for V1 and V2, maxima over random states (Appendix~\ref{app:repro}), with the threshold that the tests of \panels{} assert.}\label{tab:verification}
\footnotesize
\resizebox{\linewidth}{!}{%
\begin{tabular}{@{}llll@{}}
\toprule
Check & Configurations & Result & Test threshold \\
\midrule
V1: $\vect{c}^\top\mat{K}^{coh}_s\vect{c}$ of the kernels against the quadrature of the energy & 5 states, pristine and random $k_g$ & $\le 1.0\times10^{-15}$ & $10^{-12}$ \\
V1: matrix products, Eq.~\eqref{eq:K-update}, against the kernels & 3 states, both orders of the domains & $\le 2.7\times10^{-15}$ & $10^{-12}$ \\
V2: $\mat{K}^{coh}_s + \mat{K}^{coh}_{\dot d}$ against central finite differences & 3 states, with and without $d^h > \hat d$ & $\le 6.5\times10^{-10}$ & $10^{-6}$ \\
V2: $\mat{K}^{coh}_s$ alone against central finite differences (control) & same & 0.08--0.62 & $> 10^{-2}$ \\
V3: reaction against traction integral, linear, reference DCB & $k_o = 10^4$, $10^5$, $2\times10^5$~N/mm$^3$ & $\le 4.9\times10^{-4}$ & -- \\
V3: reaction against traction integral, nonlinear, reference DCB & cases A and B, 38 increments to 8~mm & $\le 8.9\times10^{-4}$ & -- \\
\bottomrule
\end{tabular}}
\end{table*}

\paragraph{Load (V3)} For a linear solution of the reference DCB with the reference discretisation and an opening of 0.2~mm, the reaction of Eq.~\eqref{eq:reaction} and the integral of the normal tractions with the full separation differ by $4.9\times10^{-4}$, $6.7\times10^{-8}$ and $1.4\times10^{-9}$ for $k_o = 10^4$, $10^5$ and $2\times10^5$~N/mm$^3$; the largest difference is that of the softest interface, which has the longest decay length, $1/\lambda = 2.1$~mm. In the model of the thesis, the traction integral with the corrected separation of Eq.~\eqref{eq:correction} is 14 to 28 times too high, because it drops the compressive tractions that balance the tensile ones (Appendix~\ref{app:force}). In the nonlinear analyses of the reference DCB, the reaction and the traction integral agree within $8.9\times10^{-4}$ (case~A) and $3.8\times10^{-5}$ (case~B) at every opening; the length of the cohesive domain needed for this is discussed with the convergence study (Section~\ref{sec:res-convergence}). In the validation runs, whose cohesive domains have the margin of Section~\ref{sec:dcb-models}, the two measures differ by at most 0.25\% at every opening, with two exceptions: the full weld, 7.6\% at the last opening and more than 0.5\% at the last 11 of its 60 openings, whose perfect bond beyond the cohesive domain spans the full width of the arms while the weld is 15~mm wide, and the low-strength case of Alfano and Crisfield run with $k_o = 10^5$~N/mm$^3$, 1.5\%, whose basis is too coarse for this penalty stiffness (Section~\ref{sec:res-convergence}); with $k_o = 10^3$~N/mm$^3$ the difference of this case stays below $10^{-5}$.

\paragraph{Solution procedure (V4)} With the consistent tangent, the line search and the bisection, the whole analysis of the reference DCB to 8~mm takes 134--159 iterations in 38 increments without bisection for every discretisation with $m_{tsl} \ge 15$ and $n_x \ge 120$, 3.5--4.2 per increment, and 5.5--6.0 per increment after the peak, for $m_{tsl}\geq25$ (Tables~\ref{tab:conv-grid} and~\ref{tab:conv-other}, except the halved increment). On the reference discretisation the consistent tangent alone is enough: without the line search and without bisection, every opening converges, in 2 to 7 iterations and 137 in total against 139 with the line search, which reduces the step in only 9 iterations after the peak, and the loads are the same to $10^{-4}$~N. With the secant stiffness as the only tangent of the interface, also without bisection, the iterations grow as the process zone develops, from 4--6 per increment at damage onset to 17 at $\delta = 3.7$~mm, 50 at 5.4~mm and 124 at the peak, 5.6~mm: 518 iterations up to the peak, against 71 with the consistent tangent. At the first opening after the peak the ratio of Eq.~\eqref{eq:criterion} still decreases, but only linearly, to $9\times10^{-4}$ after 200 iterations, and the run stops. The line search and the bisection are needed when the integration points are too sparse. With $n_x = 60$ (case~A, $m_{tsl} = 25$), 3.6 points per $l_{cz}$, the iterations without line search and without bisection follow those with the line search up to the peak, but at the first opening after it, $\delta = 5.8$~mm, they enter an exact cycle: from the eighth iteration on, the ratio of Eq.~\eqref{eq:criterion} repeats between 0.76 and 0.999 with a period of 12 iterations, four times the refresh interval $n_{NR} = 3$ of $\mat{K}_C + \mat{K}_G$, to better than $10^{-9}$ from the second period on, and the increment cannot converge however many iterations are allowed. With the line search, still without bisection, 251 of the 332 iterations take a reduced step, which breaks the cycle: the openings of 5.8, 6.0 and 6.2~mm converge in 8, 12 and 47 iterations, but at 6.4~mm the ratio stagnates at 0.45--0.47 with steps of $2^{-6}$ to $2^{-2}$. With the bisection as well, the same discretisation completes the analysis with 3 bisected increments and 384 iterations (Table~\ref{tab:conv-grid}). The bisection is therefore the safeguard of the coarse discretisations: 1--3 increments were bisected with $n_x = 60$ and $m_{tsl} \ge 20$, with 1.3 to 2.8 times the iterations, and more when the basis is too coarse, $m_{tsl} = 10$; it lets most of those runs complete, except for four runs that stopped at $\delta = 6.2$--7.8~mm when an increment did not converge after the last bisection (Table~\ref{tab:conv-grid}). In the 14 nonlinear runs of the validation, 800 openings in total, every opening converged without bisection in 1 to 8 iterations, 2.9 on average (Fig.~\ref{fig:runs}b).

The reference analysis to 8~mm, case~A with $m_{tsl} = 25$ and $n_x = 180$, 38 openings and 139 iterations, takes 4.6~min of wall time and 19~min of CPU time when run alone on a laptop with an AMD Ryzen~7 PRO~250 processor (16 threads, 3301~MHz, 32~GB of RAM). The validation runs, with 40--60 openings each, take 5.0--10.8~min each when run one at a time, with one thread for the linear algebra.

\subsection{Convergence of the discretisation}\label{sec:res-convergence}

\begin{figure*}[!tbp]
\centering
\includegraphics[width=\textwidth]{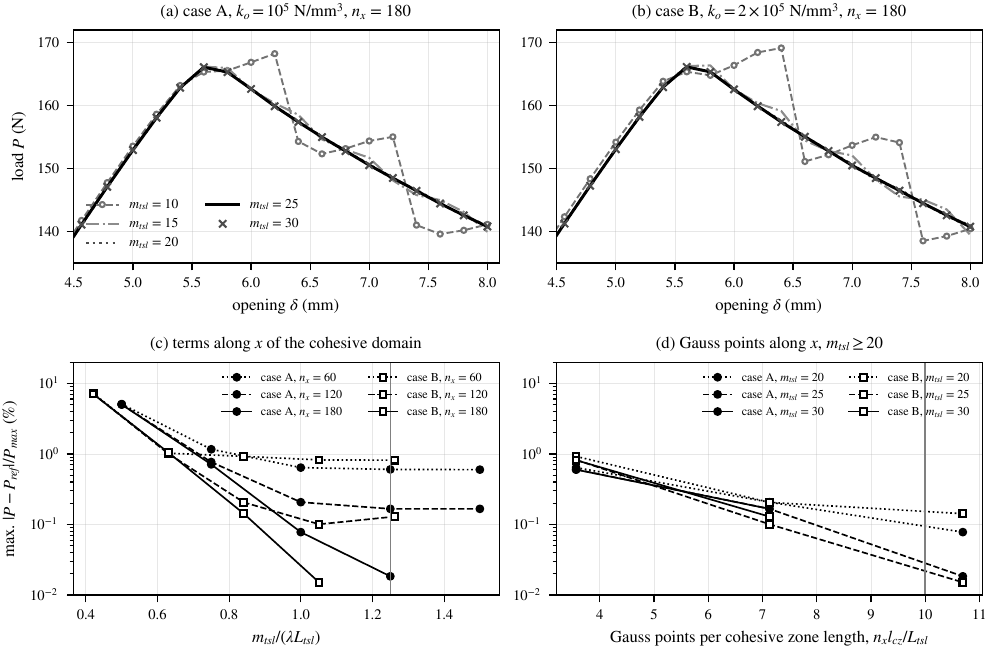}
\caption{Convergence study on the reference DCB. (a), (b) Load around and after the peak for 10 to 30 terms along $x$ of the cohesive domain, $n_x = 180$: with 10 terms the load overshoots and drops in a sawtooth, from 20 terms the curves coincide. (c) Error $e_P$ against the terms per decay length, $m_{tsl}/(\lambda L_{tsl})$, Eq.~\eqref{eq:rules}; (d) error $e_P$ against the Gauss points per cohesive zone length, $n_x l_{cz}/L_{tsl}$, for $m_{tsl} \geq 20$. Vertical lines: rules of Eq.~\eqref{eq:rules}. Reference: $m_{tsl} = 30$, $n_x = 180$.}
\label{fig:convergence}
\end{figure*}

\begin{figure*}[!tbp]
\centering
\includegraphics[width=\textwidth]{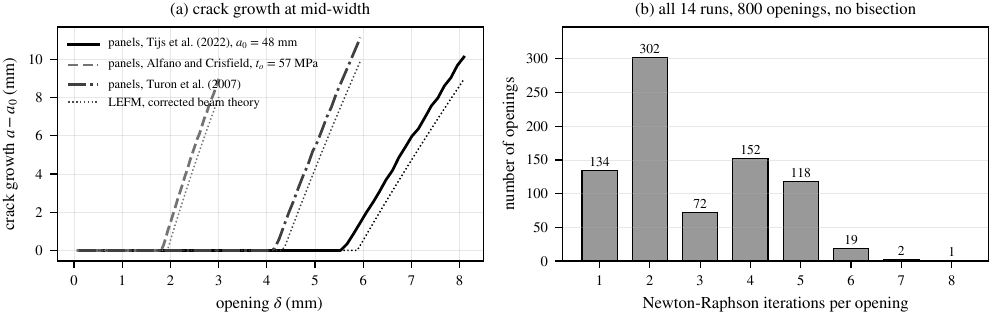}
\caption{Nonlinear runs of the validation. (a) Crack growth at mid-width, from the fully damaged points, against the crack length of the corrected beam theory at the same opening, for three unidirectional cases. (b) Newton--Raphson iterations per opening in the 14 runs.}
\label{fig:runs}
\end{figure*}

Tables~\ref{tab:conv-grid} and~\ref{tab:conv-other} and Fig.~\ref{fig:convergence} give the results of the convergence study (V5). The finest run, $m_{tsl} = 30$ and $n_x = 180$, is the reference of the error $e_P$. The wall times are those of eight runs at a time sharing the 16 threads of the laptop; the reference analysis, 10~min in this setting, takes 4.6~min when run alone (Section~\ref{sec:res-verification}).

\begin{table*}[!tbp]
\caption{Convergence study on the reference DCB, $L_{tsl} = 24$~mm: terms $m_{tsl}$ and Gauss points $n_x$ along $x$ of the cohesive domain. Loads in N, openings in mm; $e_P$: largest difference of the load to the run with $m_{tsl} = 30$, $n_x = 180$, relative to its peak load, in \%; $e_{V3}$: largest relative difference between the reaction and the traction integral (V3), in \%; time: wall time in min, eight runs at a time. The runs of case~A with $m_{tsl} = 10$ and $n_x = 60$ or 120 stopped at $\delta = 6.2$ and 7.2~mm, and those of case~B with $n_x = 60$ and $m_{tsl} = 10$ or 15 at 7.4 and 7.8~mm, when an increment did not converge after the last bisection; their $e_P$ is evaluated up to that opening. $^\ast$~Not timed: the times of these two runs were measured with an earlier version of \panels{}, with which their iterations and bisections differ.}\label{tab:conv-grid}
\footnotesize
\setlength{\tabcolsep}{3pt}
\begin{tabular*}{\tblwidth}{@{\extracolsep{\fill}}rrrrrrrrrrrrr@{}}
\toprule
$m_{tsl}$ & $n_x$ & $P_{max}$ & $\delta(P_{max})$ & $P(6)$ & $P(7)$ & $P(8)$ & $CV$ & $e_P$ & $e_{V3}$ & bisections & iterations & time \\
\midrule
\multicolumn{13}{@{}l}{case A, $k_o = 10^5$~N/mm$^3$} \\
10 & 60 & 168.36 & 6.20 & 166.88 & -- & -- & -- & 5.1 & 18 & 7 & 80 & 7 \\
10 & 120 & 168.22 & 6.20 & 166.88 & 154.38 & -- & 2.16 & 5.0 & 31 & 11 & 340 & --$^\ast$ \\
10 & 180 & 168.25 & 6.20 & 166.88 & 154.38 & 141.13 & 1.98 & 5.0 & 44 & 3 & 323 & 7 \\
15 & 60 & 166.43 & 5.60 & 162.85 & 150.95 & 139.52 & 0.63 & 1.2 & 0.27 & 3 & 283 & 8 \\
15 & 120 & 166.23 & 5.60 & 162.52 & 151.68 & 139.73 & 0.56 & 0.77 & 0.26 & 0 & 145 & 4 \\
15 & 180 & 166.19 & 5.60 & 162.53 & 151.71 & 139.95 & 0.51 & 0.71 & 0.26 & 0 & 139 & 4 \\
20 & 60 & 166.36 & 5.60 & 162.74 & 150.30 & 140.96 & 0.42 & 0.64 & 0.088 & 3 & 317 & 15 \\
20 & 120 & 166.16 & 5.60 & 162.57 & 150.63 & 140.89 & 0.26 & 0.21 & 0.088 & 0 & 143 & 6 \\
20 & 180 & 166.12 & 5.60 & 162.71 & 150.66 & 140.93 & 0.24 & 0.078 & 0.088 & 0 & 136 & 6 \\
25 & 60 & 166.35 & 5.60 & 162.68 & 150.21 & 140.78 & 0.40 & 0.60 & 0.089 & 3 & 384 & 31 \\
25 & 120 & 166.14 & 5.60 & 162.51 & 150.51 & 140.75 & 0.24 & 0.17 & 0.089 & 0 & 135 & 10 \\
25 & 180 & 166.10 & 5.60 & 162.65 & 150.54 & 140.78 & 0.23 & 0.018 & 0.089 & 0 & 139 & 10 \\
30 & 60 & 166.34 & 5.60 & 162.67 & 150.21 & 140.78 & 0.40 & 0.60 & 0.089 & 3 & 358 & 43 \\
30 & 120 & 166.14 & 5.60 & 162.51 & 150.49 & 140.77 & 0.24 & 0.17 & 0.089 & 0 & 135 & 14 \\
30 & 180 & 166.10 & 5.60 & 162.65 & 150.53 & 140.81 & 0.24 & -- & 0.088 & 0 & 140 & 16 \\
\midrule
\multicolumn{13}{@{}l}{case B, $k_o = 2 \times 10^5$~N/mm$^3$} \\
10 & 60 & 169.42 & 6.40 & 166.62 & 154.12 & -- & 2.69 & 7.2 & 18 & 7 & 140 & 8 \\
10 & 120 & 169.24 & 6.40 & 166.41 & 153.75 & 140.40 & 2.14 & 7.1 & 31 & 1 & 229 & --$^\ast$ \\
10 & 180 & 169.16 & 6.40 & 166.40 & 153.69 & 140.42 & 2.14 & 7.1 & 31 & 0 & 158 & 3 \\
15 & 60 & 166.29 & 5.80 & 162.00 & 151.84 & -- & 0.70 & 1.0 & 2.3 & 8 & 222 & 18 \\
15 & 120 & 166.36 & 5.80 & 162.54 & 152.12 & 139.47 & 0.59 & 1.0 & 2.3 & 0 & 141 & 3 \\
15 & 180 & 166.39 & 5.80 & 162.57 & 152.08 & 139.44 & 0.57 & 1.0 & 2.3 & 0 & 145 & 4 \\
20 & 60 & 166.07 & 5.60 & 162.27 & 150.90 & 142.31 & 0.51 & 0.92 & 0.032 & 2 & 238 & 11 \\
20 & 120 & 166.24 & 5.60 & 162.67 & 150.58 & 141.10 & 0.25 & 0.21 & 0.032 & 0 & 159 & 8 \\
20 & 180 & 166.20 & 5.60 & 162.70 & 150.65 & 140.94 & 0.25 & 0.14 & 0.032 & 0 & 143 & 7 \\
25 & 60 & 166.02 & 5.60 & 162.11 & 150.72 & 142.14 & 0.53 & 0.82 & 0.0059 & 1 & 189 & 13 \\
25 & 120 & 166.19 & 5.60 & 162.52 & 150.42 & 140.90 & 0.23 & 0.10 & 0.0038 & 0 & 141 & 10 \\
25 & 180 & 166.15 & 5.60 & 162.56 & 150.49 & 140.75 & 0.23 & 0.015 & 0.0038 & 0 & 140 & 10 \\
30 & 60 & 166.02 & 5.60 & 162.11 & 150.69 & 142.13 & 0.53 & 0.82 & 0.0041 & 2 & 239 & 28 \\
30 & 120 & 166.19 & 5.60 & 162.53 & 150.41 & 140.99 & 0.24 & 0.13 & 0.0039 & 0 & 137 & 15 \\
30 & 180 & 166.14 & 5.60 & 162.57 & 150.49 & 140.77 & 0.23 & -- & 0.0039 & 0 & 141 & 15 \\
\bottomrule
\end{tabular*}
\end{table*}

\begin{table*}[!tbp]
\caption{Convergence study, other parameters, varied one at a time from the reference discretisation $L_{tsl} = 24$~mm, $m_{tsl} = 25$, $n_x = 180$, $n_{tsl} = 10$, $n_y = 30$, $m = n = 8$. Same quantities and units as Table~\ref{tab:conv-grid}; $L_{tsl}$ in mm. The halved increment gives 53 openings instead of 38.}\label{tab:conv-other}
\footnotesize
\setlength{\tabcolsep}{3pt}
\begin{tabular*}{\tblwidth}{@{\extracolsep{\fill}}llrrrrrrrrrrr@{}}
\toprule
case & variation & $P_{max}$ & $\delta(P_{max})$ & $P(6)$ & $P(7)$ & $P(8)$ & $CV$ & $e_P$ & $e_{V3}$ & bisections & iterations & time \\
\midrule
A & reference & 166.10 & 5.60 & 162.65 & 150.54 & 140.78 & 0.23 & 0.018 & 0.089 & 0 & 139 & 10 \\
A & $n_{tsl} = 15$ & 166.10 & 5.60 & 162.64 & 150.54 & 140.77 & 0.23 & 0.020 & 0.089 & 0 & 136 & 25 \\
A & $n_y = 60$ & 166.10 & 5.60 & 162.63 & 150.52 & 140.77 & 0.24 & 0.021 & 0.087 & 0 & 136 & 12 \\
A & $m = n = 12$ & 166.10 & 5.60 & 162.66 & 150.54 & 140.78 & 0.24 & 0.018 & 0.087 & 0 & 134 & 13 \\
A & $\Delta\delta$ halved & 166.13 & 5.70 & 162.65 & 150.54 & 140.78 & 0.15 & 0.018 & 0.089 & 0 & 184 & 13 \\
A & $L_{tsl} = 17$ & 166.08 & 5.60 & 162.57 & 150.49 & 140.74 & 0.24 & 0.064 & 32 & 0 & 138 & 11 \\
A & $L_{tsl} = 17$, $m_{tsl} = 15$, $n_x = 120$ & 166.10 & 5.60 & 162.66 & 150.38 & 140.67 & 0.25 & 0.11 & 33 & 0 & 142 & 3 \\
\midrule
B & reference & 166.15 & 5.60 & 162.56 & 150.49 & 140.75 & 0.23 & 0.015 & 0.0038 & 0 & 140 & 10 \\
B & $n_{tsl} = 15$ & 166.14 & 5.60 & 162.56 & 150.49 & 140.73 & 0.23 & 0.028 & 0.0054 & 0 & 136 & 26 \\
B & $n_y = 60$ & 166.14 & 5.60 & 162.55 & 150.49 & 140.75 & 0.24 & 0.015 & 0.0042 & 0 & 143 & 12 \\
B & $m = n = 12$ & 166.14 & 5.60 & 162.57 & 150.49 & 140.75 & 0.23 & 0.015 & 0.0038 & 0 & 140 & 13 \\
B & $\Delta\delta$ halved & 166.21 & 5.70 & 162.56 & 150.49 & 140.75 & 0.15 & 0.015 & 0.0038 & 0 & 176 & 13 \\
B & $L_{tsl} = 17$ & 166.15 & 5.60 & 162.60 & 150.48 & 140.78 & 0.24 & 0.083 & 10 & 0 & 142 & 11 \\
B & $L_{tsl} = 17$, $m_{tsl} = 15$, $n_x = 120$ & 166.12 & 5.60 & 162.63 & 150.25 & 140.64 & 0.28 & 0.25 & 11 & 0 & 142 & 3 \\
\bottomrule
\end{tabular*}
\end{table*}

\paragraph{The Ritz basis of the cohesive domain sets the shape of the response after the peak} The opening ahead of the process zone decays with the rate $\lambda$ of Eq.~\eqref{eq:lcz}, 0.83~mm$^{-1}$ in case~A and 0.99~mm$^{-1}$ in case~B, so that the cohesive domain spans $\lambda L_{tsl} = 20.0$ and 23.8 decay lengths. With $m_{tsl} = 10$, 0.50 and 0.42 terms per decay length, the basis cannot follow the process zone: the load overshoots the converged curve up to 168.2--169.4~N at $\delta = 6.2$--6.4~mm, 1.3--2.0\% above the converged peak, and then drops by up to 14 (case~A) and 19~N (case~B) in one increment, a sawtooth with $CV = 2.0$--2.7 (Fig.~\ref{fig:convergence}a,b). The load differs from the converged one by 5--7\%, the traction integral differs from the reaction by 18--44\% from the first increment on, and three of the six runs stop at $\delta = 6.2$--7.4~mm. With $m_{tsl} = 15$ the curve is smooth, $CV = 0.51$--0.70, but still 0.7--1.2\% off. From $m_{tsl} = 20$, the error is at most 0.21\% for $n_x \ge 120$, and 25 and 30 terms give the same loads to 0.02\% with $n_x = 180$. Plotted against the terms per decay length, the errors of both cases fall on one curve, Fig.~\ref{fig:convergence}c: about 1\% at $m_{tsl} = 0.63$--$0.75\,\lambda L_{tsl}$, at most 0.14\% from $0.84\,\lambda L_{tsl}$ and below 0.1\% from $1.0\,\lambda L_{tsl}$, with $n_x = 180$. The rule $m_{tsl} \approx 1.25\,\lambda L_{tsl}$ of Eq.~\eqref{eq:rules} keeps a margin over this threshold; for $L_{tsl} = 24$~mm it asks for 25.0 terms in case~A and 29.7 in case~B, where the 25 terms of the reference discretisation, $1.05\,\lambda L_{tsl}$, are already converged, and with the cap of 25 terms it limits the cohesive domain of case~A to $25/(1.25\lambda) = 24.0$~mm.

\paragraph{The integration points set the robustness and a floor of the error} The cohesive zone length of Eq.~\eqref{eq:lcz} is $l_{cz} = 1.43$~mm in both cases, since it does not depend on $k_o$, and $n_x = 60$, 120 and 180 over 24~mm correspond to 3.6, 7.1 and 10.7 Gauss points per $l_{cz}$. With $n_x = 60$ the error stays at 0.6--0.9\% however many terms are used (Fig.~\ref{fig:convergence}c), 1 to 3 increments of every run with $m_{tsl} \ge 20$ needed bisection and the iterations increased 1.3 to 2.8 times, because the crack front crosses several integration points in one increment; the run of case~B with $m_{tsl} = 15$ stopped at $\delta = 7.8$~mm. With $n_x = 120$ the error is 0.10--0.21\% and with $n_x = 180$ at most 0.02\% (for $m_{tsl} \ge 25$), without bisections (Fig.~\ref{fig:convergence}d). This gives the second rule of Eq.~\eqref{eq:rules}, $n_x \approx 10\,L_{tsl}/l_{cz}$, 168 for the reference DCB, rounded up to 170. Unlike the terms, the integration points do not change the shape of the curve: with $n_x = 60$ the curves are smooth and the peak is within 0.15\%. The cohesive zone length enters only this second rule: the terms, which set the size of the system of equations, follow the decay rate $\lambda$, which depends on $k_o$ and $D_{11}$ but not on $\tau_o$. A short cohesive zone is therefore paid for with integration points, not with degrees of freedom, and the update of the interface stiffness, Eq.~\eqref{eq:K-update}, is restricted to the damaged points. In Table~\ref{tab:conv-grid}, 50\% more points, from $n_x = 120$ to 180 with $m_{tsl} = 25$, add less than 1~min to a wall time of 10~min, whereas 50\% more terms, from $m_{tsl} = 20$ to 30 with $n_x = 180$, more than double it. Lowering the strength to lengthen the cohesive zone, as Turon et al.~\citep{Turon2007} did to allow coarser FE meshes, is thus not needed to keep the Ritz model small, and it would bring the effect of a long cohesive zone on the peak load discussed in Section~\ref{sec:validation}.

\paragraph{Other parameters} The terms and points along $y$ and the terms of the other domains change the loads by at most 0.03\% (Table~\ref{tab:conv-other}), since the specimen is uniform across the width. Halving the increment after $\delta = 5$~mm changes only where the peak is sampled: the peak rises by 0.03~N (case~A) and 0.06~N (case~B) and moves to $\delta = 5.7$~mm, and the loads at the common openings are the same to $2\times10^{-6}$. The stiffer interface of case~B raises the initial stiffness by 1.2\%, but the peak and the loads after it differ from those of case~A by less than 0.06\%, in line with the insensitivity of the DCB to $k_o > 10^4$~N/mm$^3$ reported by Turon et al.~\citep{Turon2007}. The converged discretisation of the reference DCB is therefore $L_{tsl} = 24$~mm, $m_{tsl} = 25$, $n_x = 180$, $n_{tsl} = 10$, $n_y = 30$, with $8 \times 8$ terms in the other domains, which gives $P_{max} = 166.10$~N (case~A) and 166.15~N (case~B) at $\delta = 5.6$~mm. The validation run of the same specimen, with $L_{tsl} = 24.0$~mm, $n_x = 170$ and 60 openings to 8.1~mm (Table~\ref{tab:cases}), follows this curve within 0.17\%.

\paragraph{Length of the cohesive domain} At the last opening the crack has grown by 9.8~mm, and the margin of Section~\ref{sec:dcb-models}, $3 l_{cz} + 8/\lambda$, is 13.9~mm (case~A) and 12.4~mm (case~B), so that $L_{tsl} = 24$~mm contains the process zone and the decaying opening ahead of it up to $\delta = 8$~mm. With $L_{tsl} = 17$~mm the loads are within 0.06--0.25\% of the converged ones, but from $\delta = 7.0$ (case~A) and 7.4~mm (case~B), when the crack has grown by 6--7.5~mm, the compressive lobe enters the perfect bond, whose tractions are not part of the integral of V3, and $e_{V3}$ grows to 10--33\% at the last opening. The reaction hardly changes, so it is V3, not the load, that detects a cohesive domain that is too short; with $L_{tsl} = 24$~mm, $e_{V3}$ stays below 0.09\% (case~A) and 0.004\% (case~B), the larger value of case~A coming from its longer decay length.

\paragraph{Staircase response and discrete snap-backs} The sawtooth of Fig.~\ref{fig:convergence}a,b with 10 terms is the Ritz counterpart of the spurious oscillations of cohesive FE models with meshes that are coarse compared with the cohesive zone~\citep{Alfano2001,Turon2007}, where the Ritz basis cannot follow the process zone continuously. The Gibbs oscillations of single-domain Ritz models of progressive intralaminar damage have the same origin, a polynomial basis that cannot represent a localised damage field, and were removed by an adaptive multidomain discretisation~\citep{Campagna2024}. The elastic energy builds up while the crack front is held, and it is released when a group of integration points fails at once, a discrete snap-back that a displacement-controlled analysis crosses as a sudden load drop. Alfano and Crisfield~\citep{Alfano2001} also found that increasing the number of integration points of interface elements, which does not refine the displacement field, made their solution algorithm less robust and was not equivalent to a mesh refinement. In the Ritz model the roles are reversed in one respect, because the displacement field is refined by the number of terms, which removes the sawtooth, while the integration points discretise the damage field; more points make the stiffness change of each point smaller, remove the bisections and lower the floor of the error, but do not change the shape of the curve.

\paragraph{Discretisation of the validation runs} The validation runs use the rules of Eq.~\eqref{eq:rules} with 25 terms and the corresponding $L_{tsl} = 25/(1.25\lambda)$, i.e.\ $m_{tsl} = 1.25\,\lambda L_{tsl}$, inside the converged range of Fig.~\ref{fig:convergence}c, and with 10.0--12.3 Gauss points per $l_{cz}$ (Table~\ref{tab:cases}). The low-strength case of Alfano and Crisfield needs a cohesive domain over the whole bonded length, $L_{tsl} = 70$~mm, for which the first rule asks for 94 terms with $k_o = 10^5$~N/mm$^3$, more than the 30 Bardell functions of the current implementation. It is therefore run with $k_o = 10^3$~N/mm$^3$, for which $m_{tsl} = 30 = 1.27\,\lambda L_{tsl}$ and $n_x = 4 m_{tsl} = 120$ meet both rules. The damage onset, $\Delta_o = \tau_o/k_o = 1.7\times10^{-3}$~mm, remains far below the separation at failure, $\Delta_f = 2G_{Ic}/\tau_o = 0.33$~mm, so the softening branch that governs this case is hardly affected; Alfano and Crisfield also scaled the penalty stiffness with the strength (Table~\ref{tab:cases}). A run with $k_o = 10^5$~N/mm$^3$ and the same 30 terms, $0.4\,\lambda L_{tsl}$, shows the effect of this choice: the peaks are 50.78 and 50.81~N at the same opening, and the curves differ by at most 0.7\% of the peak, but the traction integral departs from the reaction by up to 1.5\% from the first opening with $k_o = 10^5$~N/mm$^3$, as in the runs of the convergence study with too few terms, and by less than $10^{-5}$ with $k_o = 10^3$~N/mm$^3$.

\subsection{Validation}\label{sec:validation}

\begin{figure*}[!tbp]
\centering
\includegraphics[width=\textwidth]{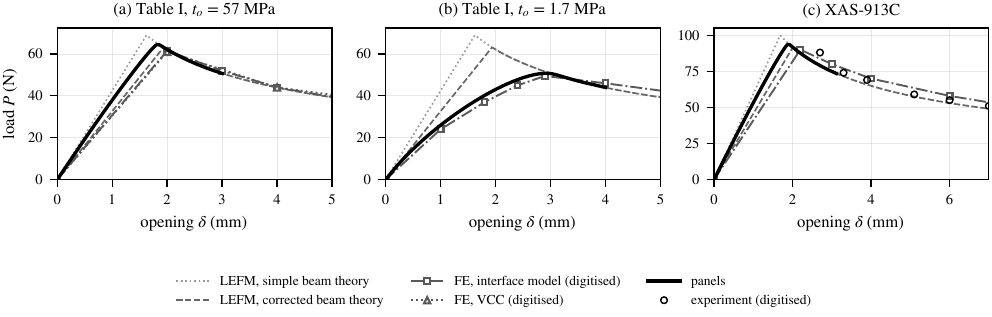}
\caption{Alfano and Crisfield~\citep{Alfano2001}: (a) DCB of Table~I with $t_o = 57$~MPa, (b) with $t_o = 1.7$~MPa (\panels{} with $k_o = 10^3$~N/mm$^3$), (c) XAS-913C DCB of Table~II. FE, interface model and virtual crack closure (VCC): digitised from Figs.~12, 15a and 23 of the reference, plotted against $\delta = 2u$; experiment: Robinson and Song~\citep{RobinsonSong1992} as plotted in Fig.~23; LEFM: Eq.~\eqref{eq:lefm}.}
\label{fig:val-alfano}
\end{figure*}

\begin{figure*}[!tbp]
\centering
\includegraphics[width=0.72\textwidth]{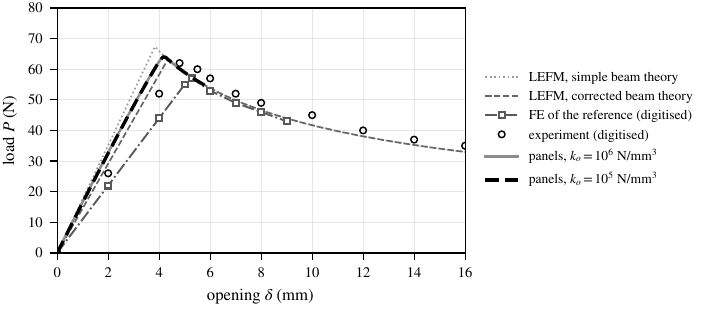}
\caption{Turon et al.~\citep{Turon2007}, T300/977-2: FE with elements of 0.125~mm and experiment digitised from Fig.~5 of the reference; \panels{} with $k_o = 10^5$ and $10^6$~N/mm$^3$. LEFM: Eq.~\eqref{eq:lefm}.}
\label{fig:val-turon}
\end{figure*}

\begin{figure*}[!tbp]
\centering
\includegraphics[width=\textwidth]{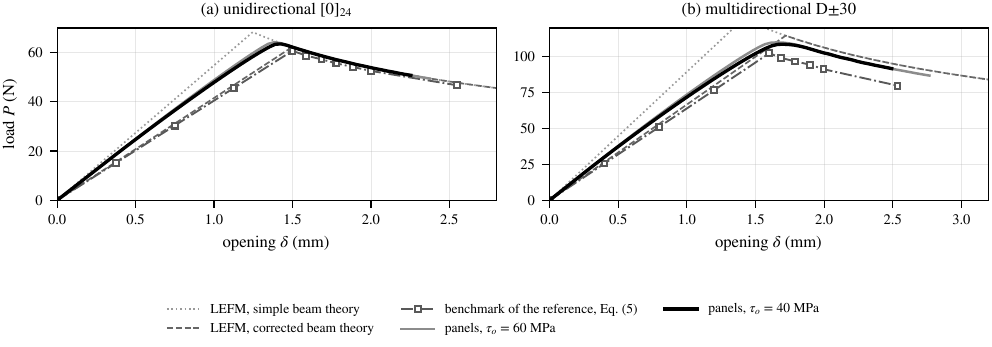}
\caption{Krueger~\citep{Krueger2008}: (a) unidirectional $[0]_{24}$ DCB of T300/1076, (b) multidirectional D$\pm$30 DCB of C12K/R6376. Benchmark: onset of delamination at mid-width, Eq.~(5) of the reference, computed from the loads of its Figs.~13 and~36 and the energy release rates of its Figs.~11 and~34. LEFM: Eq.~\eqref{eq:lefm}; \panels{} with $\tau_o = 40$ and 60~MPa.}
\label{fig:val-krueger2008}
\end{figure*}

\begin{figure*}[!tbp]
\centering
\includegraphics[width=\textwidth]{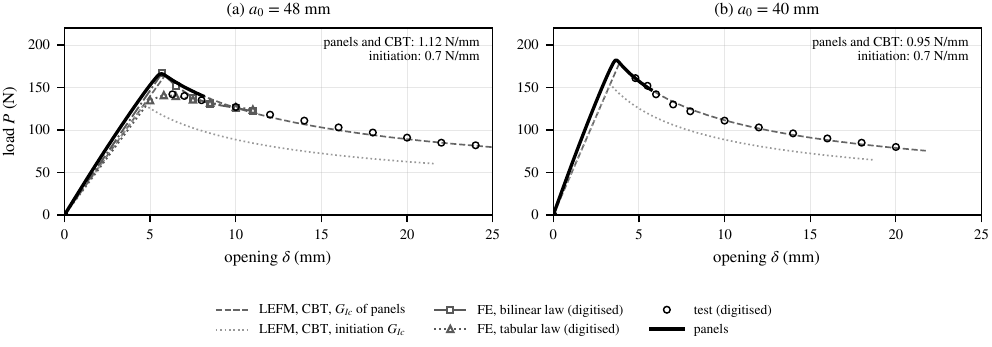}
\caption{Tijs et al.~\citep{Bas_charaterization}, AS4D/PEKK-FC: (a) $a_0 = 48$~mm, FE with the bilinear and the tabular laws digitised from Fig.~25a and test ``DCB (1)'' from Fig.~15a of the reference; (b) $a_0 = 40$~mm, test ``DCB (2)'' from Fig.~15a, \panels{} with $G_{Ic} = 0.95$~N/mm. LEFM: Eq.~\eqref{eq:lefm} with the toughness of \panels, 1.12~N/mm in (a) and 0.95~N/mm in (b), and with the initiation toughness of 0.7~N/mm.}
\label{fig:val-tijs2022}
\end{figure*}

\begin{figure*}[!tbp]
\centering
\includegraphics[width=\textwidth]{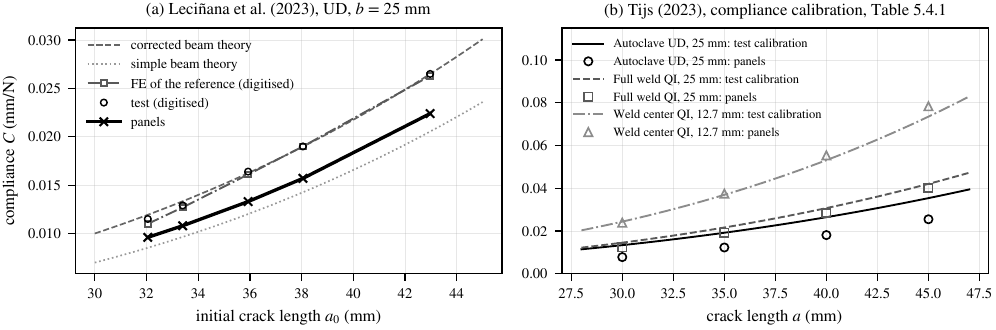}
\caption{Initial compliance. (a) Leci\~{n}ana et al.~\citep{Lecinana2023}: test and FE digitised from Fig.~10 of the reference, crack lengths of their Table~3. (b) Tijs~\citep{Tijs2023PhD}: compliance calibration $C^{1/3} = m\,a + n$ of Table~5.4.1 of the reference (lines) and \panels{} (markers). \panels: linear analyses; beam theories: Eq.~\eqref{eq:lefm}.}
\label{fig:val-compliance}
\end{figure*}

\begin{figure*}[!tbp]
\centering
\includegraphics[width=\textwidth]{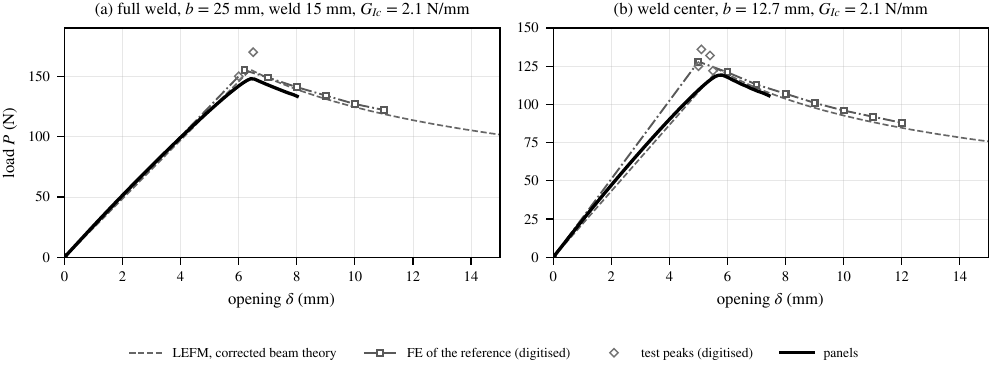}
\caption{Tijs~\citep{Tijs2023PhD}, conduction-welded quasi-isotropic DCB with $G_{Ic} = 2.1$~N/mm: (a) full weld, FE and peaks of two tests digitised from Fig.~5.5.1a; (b) weld center, FE and peaks of four tests from Fig.~5.5.1b. LEFM: Eq.~\eqref{eq:lefm} with the bonded width.}
\label{fig:val-tijs2023}
\end{figure*}

Table~\ref{tab:peaks} gives the peak loads of the 10 mode~I cases and Table~\ref{tab:compliance} the initial compliances; Figs.~\ref{fig:val-alfano}--\ref{fig:val-tijs2023} show the curves, with the data source of each curve in the legends and captions. All 14 nonlinear runs, including the sensitivity runs with $k_o = 10^6$~N/mm$^3$, with $\tau_o = 60$~MPa and, for the low-strength case of Alfano and Crisfield, with $k_o = 10^5$~N/mm$^3$, converged at every one of the 800 openings without bisection, with 4.4--11.1~mm of crack growth, and the reaction balances the integral of the cohesive tractions within 0.25\% in all of them but the two discussed with verification V3 (Section~\ref{sec:res-verification}). At the last opening, the load of \panels{} is within 0.2\% of the propagation branch of the corrected beam theory in the six unidirectional cases with nominal strengths, and the crack front, taken at the last fully damaged point at mid-width, advances at the rate predicted by LEFM, 1.0--1.2~mm ahead of the crack length of the corrected beam theory (Fig.~\ref{fig:runs}a), close to the difference of the two effective crack lengths, $\chi h - l_{cz} - 1/\lambda = 1.0$--1.4~mm.

\begin{table*}[!tbp]
\caption{Peak loads of the mode~I DCB cases. $P_{max}$ in N, $\delta$ at $P_{max}$ in mm, differences of \panels{} in \%; $\Delta K_0$: difference of the initial stiffness of \panels{} to the corrected beam theory (CBT); $\Delta a$: crack growth of the \panels{} run. $^{d}$ digitised; $^{m}$ mean of the digitised peaks of several tests (full weld: 150 and 170~N; weld center: 122--136~N); $^{b}$ not an FE curve but the VCCT benchmark of Krueger, the onset of delamination at mid-width, which lies below the load at which the width-average energy release rate reaches $G_{Ic}$ by the factors $\sqrt{G_c/\bar G} = 1.026$ (UD) and 1.123 (D$\pm$30) (see text). The two cases of Krueger are listed with $\tau_o = 40$~MPa, the reference giving no strength; with 60~MPa the peaks are 64.2 and 110.0~N. $^{k}$ $k_o = 10^3$~N/mm$^3$ (Section~\ref{sec:res-convergence}); with $10^5$~N/mm$^3$ the peak is 50.8~N at the same opening.}\label{tab:peaks}
\footnotesize
\setlength{\tabcolsep}{3pt}
\begin{tabular*}{\tblwidth}{@{\extracolsep{\fill}}lrrrrrrrrrr@{}}
\toprule
& \multicolumn{2}{c}{\panels} & \multicolumn{2}{c}{FE of the reference} & \multicolumn{2}{c}{test} & \multicolumn{2}{c}{CBT} & & \\
\cmidrule(lr){2-3}\cmidrule(lr){4-5}\cmidrule(lr){6-7}\cmidrule(lr){8-9}
Case & $P_{max}$ & $\delta$ & $P_{max}$ & \% diff. & $P_{max}$ & \% diff. & $P_{max}$ & \% diff. & $\Delta K_0$ (\%) & $\Delta a$ (mm) \\
\midrule
Alfano and Crisfield~\citep{Alfano2001}, Table~I, $t_o = 57$~MPa & 64.5 & 1.85 & 61.5$^{d}$ & +4.8 & -- & -- & 63.2 & +2.1 & +17 & 9.0 \\
Alfano and Crisfield~\citep{Alfano2001}, Table~I, $t_o = 1.7$~MPa$^{k}$ & 50.8 & 2.90 & 49.5$^{d}$ & +2.6 & -- & -- & 63.2 & $-$19.6 & $-$25 & 6.0 \\
Alfano and Crisfield~\citep{Alfano2001}, XAS-913C & 93.6 & 1.87 & 90.0$^{d}$ & +4.0 & -- & -- & 91.6 & +2.1 & +17 & 9.1 \\
Turon et al.~\citep{Turon2007}, T300/977-2 & 64.0 & 4.16 & 57.0$^{d}$ & +12.3 & 62.0$^{d}$ & +3.3 & 63.3 & +1.2 & +12 & 11.1 \\
Krueger~\citep{Krueger2008}, UD $[0]_{24}$ & 63.5 & 1.43 & 60.7$^{b}$ & +4.6 & -- & -- & 62.3 & +2.0 & +18 & 8.5 \\
Krueger~\citep{Krueger2008}, D$\pm$30 & 108.7 & 1.70 & 102.4$^{b}$ & +6.1 & -- & -- & 114.7 & $-$5.2 & +10 & 7.1 \\
Tijs et al.~\citep{Bas_charaterization}, $a_0 = 48$~mm & 166.2 & 5.66 & 167.0$^{d}$ & $-$0.5 & 142.0$^{d}$ & +17.1 & 164.5 & +1.1 & +14 & 10.1 \\
Tijs et al.~\citep{Bas_charaterization}, $a_0 = 40$~mm, 0.95~N/mm & 182.1 & 3.71 & -- & -- & 161.0$^{d}$ & +13.1 & 179.1 & +1.6 & +17 & 10.6 \\
Tijs~\citep{Tijs2023PhD}, full weld, 2.1~N/mm & 148.0 & 6.41 & 155.0$^{d}$ & $-$4.5 & 160.0$^{m}$ & $-$7.5 & 155.5 & $-$4.8 & +4 & 4.9 \\
Tijs~\citep{Tijs2023PhD}, weld center, 2.1~N/mm & 119.2 & 5.82 & 128.0$^{d}$ & $-$6.9 & 129.0$^{m}$ & $-$7.6 & 123.0 & $-$3.1 & +8 & 5.5 \\
\bottomrule
\end{tabular*}
\end{table*}

\begin{table}[!tbp]
\caption{Initial compliance: \panels{} against the measured compliances and the corrected beam theory (CBT). Differences of \panels{} in \%, over the crack lengths listed.}\label{tab:compliance}
\footnotesize
\setlength{\tabcolsep}{2.5pt}
\resizebox{\linewidth}{!}{%
\begin{tabular}{@{}lrr@{}}
\toprule
Specimens, crack lengths & vs test & vs CBT \\
\midrule
Leci\~{n}ana et al.~\citep{Lecinana2023}, Fig.~10: UD, 25~mm, & \multirow{2}{*}{$-18.9$ to $-15.5$} & \multirow{2}{*}{$-20.0$ to $-15.5$} \\
\quad $a_0 = 32.05$--42.97~mm & & \\
Tijs~\citep{Tijs2023PhD}, Table~5.4.1: autoclave UD, & \multirow{2}{*}{$-41.4$ to $-28.0$} & \multirow{2}{*}{$-20.9$ to $-14.7$} \\
\quad 25~mm, $a = 30$--45~mm & & \\
Tijs~\citep{Tijs2023PhD}, Table~5.4.1: full weld QI, & \multirow{2}{*}{$-15.2$ to $-4.7$} & \multirow{2}{*}{$-12.0$ to $-6.8$} \\
\quad 25~mm, $a = 30$--45~mm & & \\
Tijs~\citep{Tijs2023PhD}, Table~5.4.1: weld center QI, & \multirow{2}{*}{$-2.1$ to $+6.6$} & \multirow{2}{*}{$-13.5$ to $-7.5$} \\
\quad 12.7~mm, $a = 30$--45~mm & & \\
\bottomrule
\end{tabular}}
\end{table}

\paragraph{Alfano and Crisfield (2001)} With the nominal strength, \panels{} predicts 64.5~N at an opening of 1.85~mm, 4.8\% above the interface model of the paper (about 61.5~N) and 2.1\% above the corrected beam theory (63.2~N), Fig.~\ref{fig:val-alfano}a. After the peak it follows the propagation branch of LEFM: 50.7~N, 0.04\% below it, at the last opening, $\delta = 3.00$~mm.
With the low strength $t_o = 1.7$~MPa the process zone is long: the slender-body estimate $(E_{11} G_{Ic}/t_o^2)^{1/4} h^{3/4}$ of Yang and Cox~\citep{YangCox2005} gives about 15~mm, half the initial crack length, against 2.5~mm with the nominal strength. The response then departs from LEFM: the interface model peaks at about 49.5~N, at $\delta \approx 2.9$~mm, 22\% below the corrected beam theory.
\panels{} reproduces it, 50.8~N at $\delta = 2.90$~mm (+2.6\%), with the same initial stiffness (+2.3\%), Fig.~\ref{fig:val-alfano}b. This run uses $k_o = 10^3$~N/mm$^3$, for which 30 terms resolve the cohesive domain over the whole bonded length (Section~\ref{sec:res-convergence}); with $k_o = 10^5$~N/mm$^3$ and the same terms the peak is 50.8~N as well and the curves differ by at most 0.7\% of the peak. This case tests the cohesive model rather than its LEFM limit. For the XAS-913C specimen \panels{} predicts 93.6~N at 1.87~mm, 4.0\% above the interface model (about 90~N) and 2.1\% above the corrected beam theory (91.6~N), Fig.~\ref{fig:val-alfano}c. Only points after the peak were published for the experiments of Robinson and Song~\citep{RobinsonSong1992}: \panels{} is 10\% below the first, at $\delta = 2.7$~mm, and the propagation branch of the corrected beam theory, which \panels{} follows to 0.1\%, is 3.4\% below the second, at $\delta = 3.3$~mm, beyond the last opening of the run (3.12~mm), consistent with fibre bridging raising the measured propagation load.

\paragraph{Turon et al.\ (2007)} \panels{} predicts 64.0~N at 4.16~mm with $k_o = 10^5$~N/mm$^3$ and 64.1~N at 4.19~mm with the $10^6$~N/mm$^3$ of the paper, a difference of 0.2\%, in line with the insensitivity to the penalty stiffness that they report for their FE model (Fig.~\ref{fig:val-turon}). The peak is 3.3\% above the experimental peak (about 62~N) and 1.2\% above the corrected beam theory (63.3~N), and 12.3\% above their cohesive FE result with the finest mesh (about 57~N, their Figs.~5 and~8), which is 10\% below the corrected beam theory. The initial stiffness of \panels{} (16.3~N/mm) lies between the simple (17.6~N/mm) and the corrected (14.6~N/mm) beam theories, while the FE (about 11~N/mm) and experimental (about 13~N/mm) curves are 11--25\% less stiff than the corrected beam theory, more than the effect of transverse shear and root rotation that it accounts for. A possible explanation is that the plotted displacement includes compliance that no beam model has, e.g.\ of the loading hinges, or is defined differently; this is a hypothesis, which the paper does not give enough detail to confirm, and the openings at the peak are therefore not compared. The crack grew 11.1~mm with $k_o = 10^5$ and 4.4~mm with $10^6$~N/mm$^3$, whose shorter decay length $1/\lambda$ allows only a shorter cohesive domain with 25 terms.

\paragraph{Krueger (2008)} The critical load of the benchmark is the one at which $G_T$ reaches $G_c$ at mid-width, where the three-dimensional energy release rate is highest, whereas \panels{} and beam theory dissipate $G_{Ic}$ over the whole front and therefore follow its width average $\bar{G} = P_{crit}^2 a_e^2/(b\,EI)$. The two criteria differ by $\sqrt{G_c/\bar{G}}$, which is 1.026 for the unidirectional and 1.123 for the multidirectional specimen, nearly constant over the delamination lengths of the benchmark and close to the ratios of the peak of the corrected beam theory to the FE benchmark peak, $62.3/60.7$ and $114.7/102.4$. The multidirectional specimen peaks more strongly at mid-width because its more compliant arms bend anticlastically, which the reference also observes (its Section~5.3). This case therefore measures the LEFM limit of the model, and the meaningful comparison is with the corrected beam theory.

With $\tau_o = 40$~MPa, \panels{} predicts 63.5~N at 1.43~mm for the unidirectional specimen, 4.6\% above the benchmark (60.7~N at 1.50~mm) and 2.0\% above the corrected beam theory (62.3~N), and 108.7~N at 1.70~mm for the multidirectional one, 6.1\% above the benchmark (102.4~N) but 5.2\% below the corrected beam theory (114.7~N), Fig.~\ref{fig:val-krueger2008}; against the benchmark scaled by $\sqrt{G_c/\bar{G}}$, 62.3 and 115.0~N, the differences are $+2.0$\% and $-5.5$\%. The deficit of the multidirectional specimen has the same origin as in the welded joints: its cohesive zone, $l_{cz} = 1.99$~mm, is not small compared with the 2.0~mm arms, and the deficit falls to 4.1\% when $\tau_o = 60$~MPa divides $l_{cz}$ by 2.3. After the peak, the load stays within 0.1\% of the propagation branch of the corrected beam theory for the unidirectional specimen and 3.7--3.9\% below it for the multidirectional one. Since the reference gives no strength, the sensitivity to $\tau_o$ measures how much this comparison rests on a choice: from 40 to 60~MPa the peak moves by 1.1\% (63.5 to 64.2~N) and 1.2\% (108.7 to 110.0~N) and the propagation branch does not move. The strengths differ in cost instead: with $\tau_o = 60$~MPa the rule of Eq.~\eqref{eq:rules} asks for 450 Gauss columns for the unidirectional specimen, against 200 with 40~MPa. The initial stiffness is again above the corrected beam theory, by 18\% and 10\%; here the three-dimensional FE of the reference is itself only 2.5\% and 4\% less stiff than the corrected beam theory, much closer than the digitised curves of Turon et al., which supports attributing their 11--25\% difference to that figure rather than to beam theory.

\paragraph{Tijs et al.\ (2022)} For $a_0 = 48$~mm, the FE model of the paper uses the same geometry, material and bilinear law ($G_{Ic} = 1.12$~N/mm, $\tau_o = 87$~MPa) as \panels. \panels{} predicts 166.2~N at 5.66~mm against about 167~N at about 5.7~mm (their Fig.~25a), $-0.5$\% on the load and $-0.7$\% on the opening, the closest comparison of all the cases (Fig.~\ref{fig:val-tijs2022}a).
The corrected beam theory gives 164.5~N. The test peaks at about 142~N: \panels{} is 17\% above it, and the FE model of the paper with the same law 18\%.
The paper explains the difference by the R-curve of this material: the initiation toughness is about 0.7~N/mm, and the propagation toughness of 1.12~N/mm is reached only after fibre bridging develops. Their tabular multilinear law, fitted to the test, reproduces it (about 141~N), while a bilinear law with the propagation toughness cannot, which is also their conclusion. For $a_0 = 40$~mm the paper gives no FE curve, and the test ``DCB (2)'' peaks at about 161~N and then follows the LEFM curve with 0.95~N/mm, the propagation toughness without the contribution of fibre bridging (their Fig.~15a); \panels{} is run with this toughness and predicts 182.1~N, 1.6\% above the corrected beam theory with the same toughness and 13\% above the test, whose peak lies between the LEFM curves with the initiation toughness of 0.7~N/mm and with 0.95~N/mm (Fig.~\ref{fig:val-tijs2022}b). \panels{} is 9\% stiffer than the FE model of the paper and 14--17\% stiffer than the corrected beam theory, and the tests are more compliant than all models; the compliance calibration of these specimens gives a crack length correction of 9.18~mm~\citep{Tijs2023PhD}, 2.4 times the $\chi h = 3.8$~mm of the corrected beam theory.

\paragraph{Leci\~{n}ana et al.\ (2023)} \panels{} predicts initial compliances 15.5--18.9\% below the measured ones and 13--17\% below the FE results of the paper, while the corrected beam theory matches the tests within 4\%, Fig.~\ref{fig:val-compliance}a and Table~\ref{tab:compliance}. The difference is the flexibility that the CLPT arms do not have: transverse shear deformation, which is large for AS4D/PEKK-FC ($E_{11}/G_{13} = 27$), and the rotation at the crack front, which in \panels{} comes only from the elastic foundation of the cohesive interface. The corrected beam theory accounts for both through $\chi h \approx 3.8$~mm, i.e.\ an effective crack length 9--12\% longer, and a compliance about 30--40\% higher than the simple beam theory; \panels{} recovers part of it through the foundation. The same effect appears in the load--opening curves as an initial stiffness 12--18\% above the corrected beam theory for the unidirectional specimens (Table~\ref{tab:peaks}), while the peak loads, controlled by the energy release rate, stay within a few percent of it.

\paragraph{Tijs (2023): compliance calibration} For the autoclave unidirectional specimens, \panels{} is 28--41\% below the compliance calibration of the tests and 15--21\% below the corrected beam theory, Fig.~\ref{fig:val-compliance}b: as above, the CLPT arms lack the shear flexibility, and the tests contain additional compliance, since their crack length correction, 9.18~mm, is 2.4 times the $\chi h$ of the corrected beam theory. For the quasi-isotropic welded specimens, whose bending modulus (73.7~GPa) is much closer to the shear modulus, \panels{} is within 5--15\% (full weld) and 7\% (weld center) of the calibrations.

\paragraph{Tijs (2023): welded DCB} With $G_{Ic} = 2.1$~N/mm, \panels{} predicts 148.0~N for the full weld (weld of 15~mm on a 25~mm specimen) and 119.2~N for the weld center, Fig.~\ref{fig:val-tijs2023}.
These are 4.5\% and 6.9\% below the FE results of Tijs and 4.8\% and 3.1\% below the corrected beam theory.
Compared with the tests, \panels{} is 7.5\% below the mean of the two full weld tests (150 and 170~N) and 7.6\% below the mean of the four weld center tests (129~N). The cohesive zone is long, $l_{cz} = 2.7$~mm, compared with the arm thickness of 2.24~mm, which lowers the peak below the LEFM limit, as in the low-strength case of Alfano and Crisfield. The FE model of Tijs is not described in enough detail (element size, cohesive law, length of the specimen beyond the crack front) to separate this effect from its modelling assumptions. The weld narrower than the specimen is represented by the initial damage history of Section~\ref{sec:dcb-models}; the arms keep their full width.

\paragraph{Discussion} The validation supports the following conclusions.
\begin{itemize}
\item \textbf{Agreement with LEFM.} For the unidirectional specimens with nominal strengths, the peak load of \panels{} is within 2.1\% of the corrected beam theory in every case of Table~\ref{tab:peaks}, and its load after the peak within 0.2\% of the propagation branch. The model therefore dissipates $G_{Ic}$ per unit of crack area, as expected from the energetic equivalence of a cohesive law with LEFM~\citep{Alfano2001}, which also confirms that no additional energy is dissipated at crack formation (Appendix~\ref{app:ucrack}).
\item \textbf{Scatter of the FE references.} The FE results of the references scatter by about $\pm10$\% around the corrected beam theory, in both directions. The peak of Turon et al.\ with their finest mesh is 10\% below the corrected beam theory, and their FE and experimental curves are 11--25\% less stiff than it, which we attribute, as a hypothesis, to the definition of the displacement or to the compliance of the fixture.
\item \textbf{Same cohesive law.} The strongest evidence is the comparison with the FE model of Tijs et al.~\citep{Bas_charaterization}, which uses the same geometry, material and bilinear law: $-0.5$\% on the peak load and $-0.7$\% on the opening at the peak.
\item \textbf{Beyond LEFM.} With the low strength of Alfano and Crisfield, the model reproduces, within 2.6\%, an FE response 22\% below LEFM; its own peak is 19.6\% below the corrected beam theory, so that the agreement is not a consequence of the LEFM limit. A long cohesive zone also lowers the peak in the welded joints and in the multidirectional specimen of Krueger, where $l_{cz}$ is comparable with the arm thickness, and, to a lesser degree, when the strength of Krueger's specimens is lowered from 60 to 40~MPa. This is the effect that limits a reduction of the strength to coarsen a discretisation~\citep{HarperHallett2008}, where the strength changes the predicted peak, and it is kept at the value of the reference wherever the reference gives one.
\item \textbf{Experiments.} The test without a pronounced R-curve, that of Turon et al., is predicted within 3.3\%. The AS4D/PEKK-FC tests show a strong R-curve that a bilinear law with the propagation toughness cannot represent: \panels, the bilinear FE model of Tijs et al.\ and the corrected beam theory are all 11--18\% above them, and the tabular law of the paper matches them. Fibre bridging also makes the experimental curves decrease more slowly than every model after the peak.
\item \textbf{Welded joints.} \panels{} is 3.1--4.8\% below the corrected beam theory and 7.5--7.6\% below the tests; the long cohesive zone compared with the arm thickness explains part of the difference to LEFM.
\item \textbf{Pre-crack.} All the cases studied here have an initial crack. In the proposed method damage initiates at any integration point of the cohesive interface where the separation reaches $\Delta_o$, so that the same model applies to interfaces without a pre-existing crack.
\item \textbf{Plate kinematics.} The CLPT arms are too stiff for the unidirectional specimens, by 15--21\% in compliance with respect to the corrected beam theory; the peak loads are not affected, but the openings at the peak are 4--7\% smaller than those of the corrected beam theory. For the quasi-isotropic welded specimens, \panels{} is within 15\% of the measured compliance.
\end{itemize}

\subsection{Identified limitations}\label{sec:limitations}

The verification and the validation identified the following limitations of the model and of its implementation:
\begin{itemize}
\item The damage is driven by the normal separation only, and one damage variable degrades the three components. There is no mixed-mode law, which prevents the analysis of the MMB and end-notched flexure specimens and of skin--stiffener separation under general loads.
\item The current implementation considers a bilinear law and hence cannot represent R-curve effects such as fibre bridging, for which multilinear laws are needed~\citep{Bas_charaterization,Abdel_Bas_Turon_MM_MLCL}.
\item No interpenetration stiffness is used: the secant stiffness $k_o(1-d)$ also applies in compression, so a failed point offers no resistance to interpenetration. This does not affect a monotonically loaded DCB.
\item The damage history is stored once per assembly, which restricts a model to one cohesive connection.
\item The correction of the separation, Eq.~\eqref{eq:correction}, assumes a single crack front that advances along $-x$; several initiation sites require its generalisation, which is not implemented.
\item The arms follow CLPT, without transverse shear deformation: the initial compliance of unidirectional DCB specimens is 15--21\% below the corrected beam theory.
\item The Ritz basis must resolve the decay of the opening ahead of the process zone, whose rate grows with $k_o^{1/4}$, over the whole cohesive domain: with 25 terms this limits the crack growth that can be simulated to 4--11~mm in the cases studied, and a stiff interface shortens it further. When the cohesive domain must cover the whole bonded length, as in the low-strength case of Alfano and Crisfield, the penalty stiffness had to be lowered to $10^3$~N/mm$^3$ for 30 terms, the most that the current implementation of the Bardell functions in \panels{} provides, to resolve it. The convergence study indicates that $1.0\,\lambda L_{tsl}$ terms would suffice, which would lengthen the cohesive domain by 25\%, and 30 terms gave no gain over 25 on the reference DCB.
\item The analyses are displacement controlled, without arc-length control, so that snap-backs cannot be followed.
\item In the model of Section~\ref{sec:dcb-models} the perfect bond beyond the cohesive domain spans the full width of the arms, also for the narrow weld.
\end{itemize}

\section{Conclusions}\label{sec:conclusion}

A multidomain semi-analytical cohesive zone damage model was presented, verified and validated. In the multidomain method, the main structure being analysed is decomposed into multiple smaller domains for computational advantage, with each domain having its own set of continuous approximation functions based on the Ritz method. The interconnection of domains and the introduction of prescribed displacements are governed by penalty formulations. The multidomain framework is expanded to model damage, by turning the area connection of the interface into a cohesive interface whose penalty stiffness is degraded at the integration points by a bilinear traction--separation law, which tracks the initiation and the growth of damage over a two-dimensional interface. The evolution of damage, and in turn the stiffness degradation, is based on the relative normal separation at the interface, and the tangential separations use the slope of each plate at each integration point. The load is the reaction of the prescribed displacement. 

The nonlinear response was obtained with a consistent, non-symmetric tangent of the damaged interface, a linear predictor, a backtracking line search on the scaled residual norm and the bisection of increments. The consistent tangent alone follows the softening response of the reference DCB after the peak load, while the secant stiffness needs an increasing number of iterations as the process zone develops; the line search and the bisection are safeguards for discretisations with too few integration points. The interface stiffness is assembled exactly by matrix products, with an update restricted to the damaged integration points, so that the cost grows with the size of the process zone and of the cracked region and not with the size of the interface. The method verification showed that the interface stiffness and its assembly reproduce the energy of the interface to round-off, that the tangent matches central finite differences to $6.5\times10^{-10}$, and that the reaction of the prescribed displacement agrees with the integral of the normal tractions within $8.9\times10^{-4}$.
The analyses of the reference double cantilever beam (DCB) converged without bisection, in 3.5--4.2 iterations per increment on average and about six after the peak, while line search and bisection safeguard convergence for coarse discretisations.

A convergence study on 42 runs showed that the Ritz basis of the cohesive domain sets the shape of the response after the peak: too few terms give a sawtooth response that overshoots the peak by up to 2\%. The integration points set the robustness of the solution and a floor of the error, 0.6--0.9\% with 3.6 points per cohesive zone length, below 0.02\% with 10.7. This gave discretisation rules for the cohesive domain: the number of terms follows the decay of the opening ahead of the process zone, $m_{tsl} \approx 1.25\,\lambda L_{tsl}$, with the loads converged to 0.1\% from $1.0\,\lambda L_{tsl}$, and the number of integration points follows the cohesive zone length, $n_x \approx 10\,L_{tsl}/l_{cz}$. Since the cohesive zone length enters only the number of integration points, and not the number of terms that sets the size of the system, the interface strength does not have to be lowered to coarsen the model.
Such a reduction would lower the predicted peak load, as shown by the low-strength and welded cases.
The cohesive domain must sufficiently long to contain both the process zone and the decaying opening ahead of it. An insufficient length allows compressive tractions to enter the perfect bond, increasing the error between the integrated normal tractions and reaction load,  despite negligible changes in the latter. As opposed to FEM, the displacement field is refined by increasing the number of terms, removing the sawtooth oscillations, while integration points discretise the damage field. More points reduce each point's stiffness change, eliminate bisections and lower the error floor, but do not change the curve shape.

The model was validated against six references: 10 mode~I DCB cases, with finite element, virtual crack closure and experimental results, including thermoplastic and conduction-welded joints, and measured initial compliances. With the same bilinear law as a finite element model, the peak load differs by 0.5\% and the opening at the peak by 0.7\%.
For the unidirectional specimens with nominal strengths, the peak load is within 2.1\% of the corrected beam theory and the load after the peak within 0.2\% of its propagation branch; with a low interface strength, the model is 19.6\% below LEFM, within 2.6\% of the finite element reference, which is 22\% below.
The conduction-welded joints are predicted 7.5--7.6\% below the tests. The test without a pronounced R-curve is predicted within 3.3\%, and 13--17\% above those of a thermoplastic material with a strong R-curve. The significant deviation in the test case with a strong R-curve is attributed to the effects of fibre bridging, that cannot be captured by the bilinear traction separation law, used in \panels{}.
The CLPT arms underestimate the initial compliance of unidirectional specimens by 15--21\% with respect to corrected beam theory, without affecting the peak loads, because they have no transverse shear deformation and the rotation at the crack front comes only from the elastic foundation of the interface. Shear-deformable plate kinematics should mitigate this error.

Future work should address the limitations of Section~\ref{sec:limitations}: a mixed-mode law, with the validation against the MMB benchmarks of Krueger~\citep{Krueger2012} and end-notched flexure tests; multilinear laws for R-curve effects; shear-deformable kinematics; several cohesive connections per assembly; an adaptive sampling of the damage with a fixed background grid of the history; and arc-length control. These will allow skin--stiffener separation in postbuckled stiffened panels, the motivation of this work, to be analysed.

\section*{Data availability}

The model is implemented in the open-source library \panels{}~\citep{panels2026}: S.G.P.~Castro, N.~D'Souza, \textit{Semi-analytical methods for plates, shells and stiffened panels} (Version 0.14.0), Zenodo, \url{https://doi.org/10.5281/zenodo.2541522}. The source code is available at \url{https://github.com/saullocastro/panels} under the BSD-3-Clause license. The repository holds the verification tests, the DCB model, the validation notebooks with the data of the references, the scripts of the convergence study, of the solver study and of the validation runs, and the stored results read by the notebooks. The dataset of this article~\citep{DSouza2026dataset}, Zenodo, \url{https://doi.org/10.5281/zenodo.23669837}, holds what is exclusive to this article: the scripts of the figures, of the maxima of the verification, of the checks of Appendix~\ref{app:corrections} and of the statistics of the validation runs, the figures, the runs of the solver study and the timing of Appendix~\ref{app:cost}. Appendix~\ref{app:repro} lists the entry point of each result.

\section*{Acknowledgements}
This research is conducted within the research and innovation programme Luchtvaart in Transitie, which is co-funded by the Dutch National Growth Fund. Saullo G. P. Castro is supported by the ``HAPSS Phase 2'' project, and is part of The CrashProof Knowledge Centre (CrashProofLab) \cite{CrashProofLab}. Bas Tijs is supported by the ``Lichtgewicht Composieten en Constructies'' project.
The authors gratefully acknowledge Dr. Boyang Chen and Dr. ir. Frans van der Meer, both affiliated with Delft University of Technology, The Netherlands, for their invaluable guidance and insightful discussions, which contributed significantly to this work.

\section*{Declaration of competing interest}
The authors declare that they have no known competing financial interests or personal relationships that could have appeared to influence the work reported in this paper.

\printcredits

\section*{Declaration of generative AI and AI-assisted technologies in the manuscript preparation process.}
During the preparation of this work, the author(s) used Claude to cross-check all verification cases between the \panels{} repository and the manuscript, guaranteeing the claimed data reproducibility. After using this tool/service, the author(s) reviewed and edited the content as needed and take(s) full responsibility for the content of the published article.

\appendix
\counterwithin{table}{section}
\counterwithin{figure}{section}

\section{Corrections and improvements with respect to D'Souza (2024)}\label{app:corrections}

This work is based on the MSc thesis of D'Souza~\citep{DSouza2024}, which introduced the cohesive interface into the multidomain Ritz framework of \panels{}, degrading the penalty stiffness of the area connection with a traction--separation law at the integration points, and proposed the correction of the separation of Eq.~\eqref{eq:correction} (Sec.~6.3.1 of the thesis). The formulation of the thesis double-counted the dissipated energy, used an incomplete tangential separation, scaled the load with an empirical factor, and had no consistent tangent, which made the solution slow and fragile after the peak load. This appendix records, for traceability, where the formulation of Section~\ref{sec:formulation} departs from the thesis, and what the new assembly and solution procedure gain in cost over its implementation (Section~\ref{app:cost}). Section and equation numbers of the form ``Sec.~5.4'' and ``Eq.~5.38'' refer to the thesis. For each item, the problem, its consequence on the load--displacement curve of the DCB model of the thesis, and the correction are given.
That model has the specimen and the material of the reference DCB of Section~\ref{sec:dcb-models}, shortened to $L = 65$~mm, but differs from the model of that section in its domains and supports. Each arm has three domains: the cohesive domain over the whole ligament of 17~mm, a connecting domain of 0.015~mm and the free arm. The top arm is loaded at its tip by a prescribed displacement $w_p$, the tip of the bottom arm is simply supported, and the rotation of the far edge of the top cohesive domain is constrained.
 Table~\ref{tab:corrections} summarises the corrections and indicates whether the problem was in the text of the thesis, in its implementation, or in both. The corrections are part of \panels{} since PR~\#15, and therefore of the version 0.14.0 used in this work (Appendix~\ref{app:repro}).

\begin{table*}[!tbp]
\caption{Summary of the corrections to the formulation and the implementation of D'Souza~\citep{DSouza2024}.}\label{tab:corrections}
\footnotesize
\begin{tabularx}{\textwidth}{@{}lp{0.2\textwidth}lXl@{}}
\toprule
Item & Location in the thesis & In & Consequence & Correction \\
\midrule
\ref{app:ucrack} Energy of crack formation & Secs.~5.1--5.2, Eqs.~5.13, 5.18, 5.26, 5.28, 5.29 & text, code & dissipation counted twice; sharp stiffening at damage onset; a term of the tangent dropped to converge & Eq.~\eqref{eq:residual} \\
\ref{app:force} Load measurement & Sec.~5.4, Eq.~5.38, Sec.~7.1.4 & text, code & compressive tractions dropped; empirical scaling fixed at the first increment & Eq.~\eqref{eq:reaction} \\
\ref{app:edge} Edge penalties & DCB use case of Sec.~7 & code & connections at the crack front act as hinges; elastic stiffness 20\% low & $100 \times$ defaults \\
\ref{app:kinematics} Tangential separation & Eqs.~4.57--4.62, 5.1 & text, code & spurious tangential separation in the process zone & Eq.~\eqref{eq:jump} \\
\ref{app:secant} Secant stiffness of the interface & Sec.~5.3 & code & residual inconsistent with the law between tangent updates & Eq.~\eqref{eq:residual} \\
\ref{app:correction-impl} Correction of the separation & Sec.~6.3.1 & code & whole rows, including the process zone, zeroed & Eq.~\eqref{eq:correction} \\
\ref{app:criterion} Convergence criterion & Eq.~5.35 & text, code & tolerance relative to the penalty terms, not to the physical forces & Eq.~\eqref{eq:criterion} \\
\ref{app:cost} Cost of the interface assembly & implementation & code & loops over all terms and points at every evaluation; 67\% of the run time in the elastic range & Eqs.~\eqref{eq:K-products}--\eqref{eq:K-update} \\
\bottomrule
\end{tabularx}
\end{table*}

\subsection{Energy of crack formation}\label{app:ucrack}

The thesis adds to the total potential energy, Eq.~5.29, the energy dissipated by crack formation, $U_{crack} = \int_A G \dd A$ (Eq.~5.18), with $G$ computed from the stiffness and the separation of the last two states (Eq.~5.13), the force vector $\vect{f}_{crack}$ of Eq.~5.28 and the stiffness matrix $\mat{K}_{crack} = \mat{K}^1_{crack} - \mat{K}^2_{crack}$ of Eq.~5.26. This counts the dissipated energy twice. The degradation of the penalty stiffness $k^w_{CZ} = k_o(1-d)$ inside $\mat{K}^{conn}_{dmg}$, Eq.~5.5, already makes the traction follow the softening branch of the law, and the work of the traction on a point up to its failure is the area under the branch, $G_{Ic}$, as in an FE model with a cohesive law. The sudden increase of the slope of the load--displacement curve at damage onset reported in Sec.~7.1.4 of the thesis, and the need to drop $\mat{K}^2_{crack}$ to obtain convergence, are consequences of the double counting. The formulation of Section~\ref{sec:cohesive} has neither $U_{crack}$, $\vect{f}_{crack}$ nor $\mat{K}_{crack}$, and the residual is Eq.~\eqref{eq:residual}. The agreement of the load after the peak with the propagation branch of LEFM within 0.2\% (Section~\ref{sec:validation}) confirms that no energy is missing.

\subsection{Load measurement}\label{app:force}

The thesis evaluates the load with two approaches (Sec.~5.4): the line integral of the transverse shear force $Q_x$ at the loaded edge (Eq.~5.37), with $Q_x$ from finite differences of the moments, and the area integral of the traction (Eq.~5.38). In the implementation, the second was computed with the corrected separation of Sec.~6.3.1, in which the separation beyond the first non-positive point is set to zero. This removes the compressive tractions that balance the tensile ones: for a beam on an elastic foundation the separation behind the peak is physically negative, and its integral is not small. The thesis then scaled the second approach by the ratio between both at the first increment (Sec.~7.1.4). Table~\ref{tab:force} compares, for a linear solution with a tip displacement of 1~mm, the reaction of the prescribed displacement with the integrals of the traction with the full separation $\Delta$ and with the corrected separation $\Delta^{corr}$. The ratio between the corrected integral and the reaction, 14 to 28, depends on $k_o$ and on the damage state, so that a factor fixed at the first increment distorts the curve after damage onset; the line integral of $Q_x$ is 2\% low. The load is now the reaction, Eq.~\eqref{eq:reaction}, and the traction integral with the full separation is used as a check of equilibrium (verification V3).

\begin{table}[!tbp]
\caption{Out-of-plane force of the DCB model of the thesis for a tip displacement of 1~mm, linear solution, default edge penalties. Forces in N.}\label{tab:force}
\footnotesize
\setlength{\tabcolsep}{3pt}
\begin{tabular*}{\tblwidth}{@{\extracolsep{\fill}}rrrrr@{}}
\toprule
$k_o$ (N/mm$^3$) & Reaction & $\int k \Delta \dd A$ & $\int k \Delta^{corr} \dd A$ & $\int Q_x \dd y$ \\
\midrule
$1\times10^4$ & 25.30 & 25.31 & 353.0 & 24.73 \\
$5\times10^4$ & 26.18 & 26.18 & 523.9 & 25.59 \\
$2\times10^5$ & 26.69 & 26.70 & 741.2 & 26.09 \\
\bottomrule
\end{tabular*}
\end{table}

\subsection{Penalty stiffness of the edge connections}\label{app:edge}

The default penalties of the edge connections, $k_t = A_{11}/h$ and $k_r = D_{11}/h$, were used in the thesis for equal laminates~\citep{Saullo_MD_Delam}. The rotational penalty per unit width of Eq.~\eqref{eq:pen-edge} is a spring of stiffness $k_r$, which adds to a cantilever of length $a$ loaded at its tip a compliance equal to $3h/a$ of its bending compliance, about 13\% for the arm of the reference DCB ($h = 2.1$~mm, $a = 48$~mm). The arms of the model of the thesis have two connections each at the crack front, where the moment is largest. Table~\ref{tab:edge} gives the elastic stiffness of the DCB with $k_o = 2\times10^5$~N/mm$^3$ when the default penalties are multiplied by a factor. The converged value agrees with two beams on an elastic foundation~\citep{Kanninen1973}, $P/\delta = 33.0$~N/mm. With the default penalties, the stiffness is 20\% low, and the separation at the crack front, and therefore the damage onset, lags the imposed displacement by 21\%. This does not depend on the connecting domain of 0.015~mm: replacing it by a domain of 5~mm gives 27.23~N/mm with the default penalties and 33.28~N/mm with the factor 100. The DCB use case of the thesis and the model of Section~\ref{sec:dcb-models} multiply the default penalties by 100; the defaults of the \panels{} library are unchanged.

\begin{table}[!tbp]
\caption{Elastic stiffness of the DCB model of the thesis ($k_o = 2\times10^5$~N/mm$^3$) against the factor applied to the default edge penalties, and maximum separation of the interface per unit of imposed displacement.}\label{tab:edge}
\footnotesize
\begin{tabular*}{\tblwidth}{@{\extracolsep{\fill}}rrr@{}}
\toprule
Factor & $P/\delta$ (N/mm) & $\Delta_{max}/\delta$ \\
\midrule
1 & 26.69 & $4.22\times10^{-4}$ \\
10 & 32.54 & $5.21\times10^{-4}$ \\
100 & 33.27 & $5.35\times10^{-4}$ \\
1000 & 33.35 & $5.36\times10^{-4}$ \\
\bottomrule
\end{tabular*}
\end{table}

\subsection{Tangential separation of the cohesive interface}\label{app:kinematics}

The compatibility of the area connection, thesis' Eqs.~4.57--4.62, assumes equal slopes of the two domains, $w^i_{,x} = w^j_{,x}$ and $w^i_{,y} = w^j_{,y}$ (Eq.~4.61), to write the in-plane compatibility with the slope of the top domain only (Eq.~4.62), which Eq.~5.1 uses for the damaged interface:
\begin{equation}\label{eq:thesis-jump}
\Delta u^{thesis} = u^t + (d^t + d^b)\, w^t_{,x} - u^b.
\end{equation}
This holds only for a perfect bond, but not inside the fracture process zone, where the two arms rotate in opposite directions. There, Eq.~\eqref{eq:thesis-jump} gives a spurious tangential separation, penalised with the same stiffness $k^w_{CZ}$, which stiffens the interface exactly where it should open. The interface now uses the jump of Eq.~\eqref{eq:jump}, with the slope of each domain. With $f$, $g$ the approximation functions of Eq.~\eqref{eq:ritz}, $w_{,x} = (2/a_p) f_{,\xi}\, g$ and $w_{,y} = (2/b_p) f\, g_{,\eta}$, the terms of the kernel that are new or modified with respect to the thesis are, before the factor $k_g w_\xi w_\eta\, a_p b_p/4$ of the quadrature (subscripts $u$, $v$, $w$ indicate the functions of each displacement, superscripts the domain):
\begingroup
\binoppenalty=10 \relpenalty=10 \emergencystretch=2em
\begin{itemize}
\item block $tt$: the terms of the thesis with $d^t$ in place of $d^t + d^b$;
\item block $tb$: $(u^t, w^b)$: $+\frac{2}{a_p} d^b f^t_u f^b_{w,\xi} g^t_u g^b_w$; $(v^t, w^b)$: $+\frac{2}{b_p} d^b f^t_v f^b_w g^t_v g^b_{w,\eta}$; $(w^t, u^b)$ and $(w^t, v^b)$ with $d^t$; $(w^t, w^b)$: $-f^t_w f^b_w g^t_w g^b_w + \frac{4 d^t d^b}{a_p^2} f^t_{w,\xi} f^b_{w,\xi} g^t_w g^b_w + \frac{4 d^t d^b}{b_p^2} f^t_w f^b_w g^t_{w,\eta} g^b_{w,\eta}$;
\item block $bb$: $(u^b, w^b)$: $-\frac{2}{a_p} d^b f_u f_{w,\xi} g_u g_w$; $(v^b, w^b)$: $-\frac{2}{b_p} d^b f_v f_w g_v g_{w,\eta}$; $(w^b, w^b)$: $+\frac{4 (d^b)^2}{a_p^2} f_{w,\xi} f_{w,\xi} g_w g_w + \frac{4 (d^b)^2}{b_p^2} f_w f_w g_{w,\eta} g_{w,\eta}$.
\end{itemize}
\endgroup
These blocks were verified by V1 (Table~\ref{tab:verification}). The kernel of the thesis also built the functions along $\eta$ of the top domain in the block $tb$ with the flags of the edge $y_1$ in place of $y_2$, an error of the implementation that had no effect on the DCB, where all flags along $y$ are 1. The kernel of the perfect bond keeps Eq.~\eqref{eq:thesis-jump}, which is adequate for it.

\subsection{Secant stiffness of the interface in the modified Newton--Raphson method}\label{app:secant}

In the modified Newton--Raphson method of the thesis in Sec.~5.3, the implementation refreshed $\mat{K}^{conn}_{dmg}$ only every $n_{NR} = 3$ iterations, together with the tangent of the domains. However, $\mat{K}^{conn}_{dmg}$ is the secant stiffness of the interface and part of the internal force of the residual (Eq.~5.32). Keeping it frozen makes the residual inconsistent with the traction--separation law, and a converged state may carry a damage field computed with the displacements of earlier iterations. The secant stiffness is now evaluated at every evaluation of the residual, and only $\mat{K}_C + \mat{K}_G$ is refreshed every $n_{NR}$ iterations. The iteration matrix of the thesis also lacked the damage-rate term of Eq.~\eqref{eq:tangent}, which the thesis did not derive.

\subsection{Correction of the separation}\label{app:correction-impl}

Section~6.3.1 of the thesis describes the correction of the separation as in Section~\ref{sec:cohesive}: from the maximum separation, location $A$, move in the direction of decreasing separation up to the first negative separation, location $B$, and set the separation to zero beyond $B$. The implementation set to zero, for each row at constant $y$, everything from $x = 0$ up to the last non-positive separation of the row. When the approximation functions give a non-positive value between $A$ and the edge of the domain at the crack front, for instance a small oscillation at that edge, the whole row, including the process zone, was set to zero. The implementation now follows the description of the thesis, Eq.~\eqref{eq:correction}; the text of the thesis was correct. The negative separations are not only numerical: for a beam on an elastic foundation the separation ahead of the process zone oscillates with a decaying amplitude, and its first negative lobe does not change when the number of terms of the cohesive domain is increased from 10 to 20.

\subsection{Convergence criterion}\label{app:criterion}

The prescribed displacement is imposed with $k_w = 10^6$~N/mm$^2$, such that $\vect{f}_{PD}$ and $\mat{K}^{pen}_{PD}\vect{c}$ are of the order of $10^6$--$10^7$, while the physical forces are of the order of $10^2$~N. The denominator of the criterion of the thesis, $\max(F(\vect{f}^{net}_{ext}), F(\vect{f}^{net}_{int}))$ (Eq.~5.35), is dominated by these penalty terms, and the tolerance is then not relative to the physical forces. The criterion now uses the internal force and the reaction of the prescribed displacement, Eq.~\eqref{eq:criterion}. The large $k_w$ also degrades the first solution of each increment, whose ratio can be close to 1; the analysis is therefore stopped only when the ratio diverges, and not when it reaches 1 as in the implementation of the thesis.

\subsection{Cost of the interface assembly}\label{app:cost}

The implementation of the thesis assembled the stiffness of the cohesive interface with loops over the terms of both domains and over the integration points, evaluating the approximation functions inside the loops, at every evaluation of the residual. For the DCB model of the thesis, with $60 \times 30$ integration points and $10 \times 10$ and $8 \times 8$ terms, one assembly took 6.5~s, whatever the damage, and the 22 assemblies of the eight elastic increments took 67\% of the run time. With the matrix products of Eqs.~\eqref{eq:K-products}--\eqref{eq:K-update}, the assembly of the same interface takes 0.038~s for the pristine interface and 0.05--0.08~s when 20--100\% of the points are damaged, i.e.\ 85--170 times faster. The kernels that loop over the terms and the points, kept for domains with different planforms (Section~\ref{sec:assembly}), now evaluate the approximation functions once per integration point and take 0.49~s, 13 times faster than the loops of the thesis but still 7--13 times slower than the matrix products. Together with the predictor of Eq.~\eqref{eq:predictor} and the reuse of the tangent of the domains, the eight elastic increments of this model take 7.1~s instead of 218~s with the code of the thesis. Without the predictor and the reuse, and with the kernels, they take 24~s, with the same loads to a relative $3.0\times10^{-8}$, the order of the convergence tolerance. All times are wall times measured on the laptop of Section~\ref{sec:res-verification}, one analysis at a time, and are the medians of three runs, between which they varied by up to 50\%.

\section{Reproducibility map}\label{app:repro}

Table~\ref{tab:repro} gives, for each result of Sections~\ref{sec:res-verification}--\ref{sec:validation} and of Appendix~\ref{app:corrections}, the entry point, how to run it and the stored data used.

\paragraph{Common set-up} The results were obtained with \panels{} version 0.14.0~\citep{panels2026}, \texttt{structsolve} 0.9.3 and \texttt{composites} 0.10.1, with \texttt{conn\_method='penalty'}. The repository holds the tests, the DCB model of Section~\ref{sec:dcb-models} (\texttt{notebooks/dcb\_utils.py}), the validation notebooks with the data of the references, the scripts of the convergence study, of the solver study and of the validation runs, and the stored results read by the notebooks; \texttt{notebooks/README.md} lists them. The dataset of this article~\citep{DSouza2026dataset} holds only what the article needs and the repository does not store: the scripts of the figures, of the maxima of the verification, of the checks of Appendix~\ref{app:corrections} and of the statistics of the validation runs, the figures, the runs of the solver study and the timing of Appendix~\ref{app:cost}; its \texttt{README.md} lists them. The notebooks import \texttt{dcb\_utils.py} and must be run from the folder \texttt{notebooks/}. By default (\texttt{RERUN = False}) they load the stored \panels{} curves of \texttt{notebooks/results/*.npz}, read with \texttt{dcb\_utils.load}: the curves (\texttt{delta}, \texttt{P}, \texttt{P\_traction}, \texttt{a}, \texttt{dmax}), the iterations (\texttt{iters}, \texttt{failed}), the wall and CPU times, the model and its discretisation (\texttt{case}), the solver settings (\texttt{solver}), the version of \panels{} and the connection method; with \texttt{RERUN = True} they run the nonlinear analyses again, one after the other, and an incomplete results file is completed automatically. The environment variable \texttt{DCB\_CASES} restricts the notebooks of Krueger, Tijs et al., Leci\~{n}ana et al.\ and Tijs to some cases. The linear compliance analyses are not stored and take seconds. The data of the references are typed into the dictionaries of the notebooks, which state their source, the digitisation uncertainty and the assumptions. Wall times depend on the hardware: they were measured on a laptop with an AMD Ryzen~7 PRO~250 processor (16 threads, 32~GB of RAM), with eight runs at a time for the convergence study, and one run at a time for the validation, the reference analysis of V4 and Appendix~\ref{app:cost}. The run times of the convergence study were measured with \panels{} 0.13.11, \texttt{structsolve} 0.8.0 and \texttt{composites} 0.9.21, when the study was first computed; they should be equivalent for \panels{} 0.14.0, since no change aimed at performance was made between these versions.
\begin{table*}[!tp]
\caption{Reproducibility map. \texttt{PANELS}: the \panels{} repository at the tag 0.14.0, built in place; paths relative to its root: \texttt{tst/} = \texttt{tests/multidomain/}, \texttt{nb/} = \texttt{notebooks/}, \texttt{res/} = \texttt{notebooks/results/}, \texttt{cs.py} = \texttt{nb/dcb\_mode\_I\_convergence\_study.py} and \texttt{ss.py} = \texttt{nb/dcb\_mode\_I\_solver\_study.py}, run from the root; the notebooks are run from \texttt{nb/}. \texttt{ds/}: the dataset of this article~\citep{DSouza2026dataset}; its scripts, in \texttt{ds/scripts/}, are run from that folder.}\label{tab:repro}
\scriptsize
\setlength{\tabcolsep}{3pt}
\renewcommand{\arraystretch}{1.0}
\begin{tabularx}{\textwidth}{@{}>{\raggedright\arraybackslash}p{0.13\textwidth}>{\raggedright\arraybackslash}p{0.17\textwidth}>{\raggedright\arraybackslash}p{0.19\textwidth}>{\raggedright\arraybackslash}p{0.14\textwidth}>{\raggedright\arraybackslash}X@{}}
\toprule
Manuscript item & Entry point & How to run & Stored data & Remarks \\
\midrule
Table~\ref{tab:verification}, V1 and V2 & \texttt{ds/scripts/verification\_v1\_v2.py} & \texttt{python verification\_v1\_v2.py --repo PANELS} & none & maxima over the random states of the table: seeds 1--5 (energy), 5--7 (matrix products), 3, 4 and 8 (tangent); the tests \texttt{tst/test\_sb\_tsl.py} (\texttt{pytest -k "energy or matrix\_products or finite\_differences"}) assert the thresholds for one state each \\
Table~\ref{tab:verification}, V3 & \texttt{ds/scripts/verification\_v3.py} & \texttt{python verification\_v3.py --repo PANELS} & \texttt{res/dcb\_mode\_I\_convergence\_study.json} & linear: reference DCB and discretisation of \texttt{cs.py} (\texttt{DCB}, \texttt{REF}), one opening of 0.2~mm, $k_o = 10^4$, $10^5$, $2\times10^5$~N/mm$^3$; nonlinear: runs \texttt{A\_m25\_nx180} and \texttt{B\_m25\_nx180} \\
Tables~\ref{tab:conv-grid}, \ref{tab:conv-other}, Fig.~\ref{fig:convergence}; iterations and bisections of V4 & \texttt{cs.py} & \texttt{python cs.py run --jobs 8 --longest-first res/dcb\_mode\_I\_convergence\_study.json}, then \texttt{python cs.py collect --json FILE} & \texttt{res/dcb\_mode\_I\_convergence\_study.json} & 42 runs of \texttt{nb/dcb\_utils.py}: \texttt{build\_dcb}, \texttt{solve\_dcb}; cases A and B; keys \texttt{runs/TAG/delta}, \texttt{P}, \texttt{P\_traction}, \texttt{iters}, \texttt{bisections}, \texttt{wall\_min}, \texttt{cpu\_min}; wall and CPU times of the first computation with \panels{} 0.13.11 \\
V4: secant tangent, no line search, $n_x = 60$; wall time of the reference analysis & \texttt{ss.py} & \texttt{python ss.py run ref} alone, then \texttt{run secant nols nols\_nx60 ls\_nx60 --jobs 4} and \texttt{collect} & \texttt{ds/solver\_study\_runs/} & reference DCB, case~A; without bisection, up to 200 iterations per opening; every Newton--Raphson iteration traced in \texttt{TAG\_trace.npy}; \texttt{ss.py} writes to \texttt{PANELS/dcb\_mode\_I\_solver\_study\_runs/} \\
Nonlinear runs of the validation & \texttt{nb/dcb\_mode\_I\_validation\_rerun.py} & \texttt{python nb/dcb\_mode\_I\_validation\_rerun.py run --jobs 7} & \texttt{res/*.npz} & records the runs of the notebooks below and solves them in parallel, one process per run, overwriting their results; 17 runs, of which the 14 of this article \\
Section~\ref{sec:validation} and V3, V4: openings, iterations, crack growth, wall times and V3 of the validation runs; last opening against CBT; Fig.~\ref{fig:runs} & \texttt{ds/scripts/validation\_statistics.py} & \texttt{python validation\_statistics.py --repo PANELS} & \texttt{res/*.npz}, \texttt{res/dcb\_mode\_I\_convergence\_study.json} & keys \texttt{P}, \texttt{P\_traction}, \texttt{a}, \texttt{iters}, \texttt{failed}, \texttt{time}; CBT with \texttt{corrected=True}; also the validation run of the reference DCB against the converged run \\
Table~\ref{tab:peaks} & \texttt{nb/dcb\_mode\_I\_validation\_summary.ipynb} & run all cells & all \texttt{res/*.npz} of the validation & collects the notebooks below, with the sensitivity runs to $k_o$ and $\tau_o$; its rows of the weld center with 1.95 and 1.12~N/mm and of Camanho et al.\ are not used in this article \\
Fig.~\ref{fig:val-alfano} & \texttt{nb/dcb\_alfano2001.ipynb} & run all cells & \texttt{res/alfano2001\_TableI\_to57.npz}, \texttt{\_TableI\_to1.7.npz}, \texttt{\_TableI\_to1.7\_K1e5.npz}, \texttt{\_TableII\_XAS913C.npz} & $\delta = 2u$; low-strength case with $k_o = 10^3$ and, for comparison, $10^5$~N/mm$^3$ \\
Fig.~\ref{fig:val-turon} & \texttt{nb/dcb\_turon2007.ipynb} & run all cells & \texttt{res/turon2007\_le0.125\_tau60.npz}, \texttt{\_K1e6.npz} & $k_o = 10^5$ and $10^6$~N/mm$^3$ \\
Fig.~\ref{fig:val-krueger2008} & \texttt{nb/dcb\_krueger2008.ipynb} & run all cells & \texttt{res/krueger2008\_DCB\_UD\_tau40.npz}, \texttt{\_UD\_tau60.npz}, \texttt{\_MD\_tau40.npz}, \texttt{\_MD\_tau60.npz} & the benchmark of Eq.~(5) of the reference is derived in the notebook from the digitised Figs.~11, 13, 34 and 36; $\tau_o$ is a choice \\
Fig.~\ref{fig:val-tijs2022} & \texttt{nb/dcb\_tijs2022.ipynb} & run all cells & \texttt{res/tijs2022\_a48.npz}, \texttt{\_a40\_G0.95.npz} & $G_{Ic} = 1.12$~N/mm ($a_0 = 48$~mm) and 0.95~N/mm ($a_0 = 40$~mm) \\
Table~\ref{tab:compliance}, Fig.~\ref{fig:val-compliance}a & \texttt{nb/dcb\_lecinana2023.ipynb} & run all cells & none & linear analyses, \texttt{dcb\_utils.elastic\_compliance} \\
Table~\ref{tab:compliance}, Figs.~\ref{fig:val-compliance}b and \ref{fig:val-tijs2023} & \texttt{nb/dcb\_tijs2023\_phd.ipynb} & run all cells & \texttt{res/tijs2023\_full\_weld.npz}, \texttt{\_weld\_center.npz} & elastic properties and specimen length assumed; the notebook also runs the weld center with $G_{Ic} = 1.95$ and 1.12~N/mm, not used in this article \\
Figs.~\ref{fig:convergence}--\ref{fig:val-tijs2023} & \texttt{ds/scripts/make\_figures.py} & \texttt{python make\_figures.py --repo PANELS} & as in the rows above & reads the stored results, and the data of the references from the notebooks, executed without running any analysis; writes \texttt{ds/figs/} \\
Figs.~\ref{fig:tsl}, \ref{fig:flowchart} and \ref{fig:dcb-model} & \texttt{ds/scripts/make\_schematics.py} & \texttt{python make\_schematics.py} & none & schematic figures \\
Eqs.~\eqref{eq:lefm}--\eqref{eq:chi} & \texttt{nb/dcb\_utils.py}: \texttt{lefm\_curve}, \texttt{compliance\_cbt}, \texttt{chi\_williams} & called by the notebooks & none & $G_{13} = G_{12}$; bending modulus $12D_{11}/h^3$ for the multidirectional and quasi-isotropic arms \\
Eq.~\eqref{eq:rules}, Table~\ref{tab:cases} ($L_{tsl}$, $m_{tsl}$, $n_x$) & \texttt{nb/dcb\_utils.py}: \texttt{resolution}, \texttt{discretize}, \texttt{margin}, \texttt{opening\_at\_crack\_length} & called by \texttt{setup} in each notebook with nonlinear runs & none & the printed output of these notebooks gives $\lambda$, $l_{cz}$, $L_{tsl}$, $m_{tsl}$, $n_x$ and the last opening \\
Tables~\ref{tab:force} and~\ref{tab:edge}; Appendices~\ref{app:edge} and~\ref{app:correction-impl} & \texttt{ds/scripts/thesis\_model\_checks.py} & \texttt{git archive 0b619a1} of \texttt{PANELS} into a folder THESIS, built in place, then \texttt{python thesis\_model\_checks.py --thesis-root THESIS} & none & linear solutions of the model of the thesis (\texttt{tst/test\_dcb\_damage.py}: \texttt{dcb\_damage\_prop\_no\_f\_kcrack}) with its code, \panels{} 0.6.21, commit \texttt{0b619a1}; also the lag of the separation, the connecting domain of 5~mm, two beams on an elastic foundation and the negative separation with 10 and 20 terms \\
Appendix~\ref{app:cost}, cost of the assembly and of the elastic range & \texttt{ds/scripts/thesis\_timing.py} & THESIS as above, then \texttt{python thesis\_timing.py run --repeat 3 --repo PANELS --thesis-root THESIS}, alone, and \texttt{summary} & \texttt{ds/scripts/thesis\_timing\_runs/repetition\_*/*.json} & model of the thesis, eight elastic increments; current code with the matrix products, with the kernels (\texttt{use\_kernels=True}, \texttt{predictor=False}, \texttt{kT\_pan\_reuse\_steps=0}), and the code of the thesis; then single assemblies with 0--100\% of the points damaged \\
\bottomrule
\end{tabularx}
\end{table*}

\bibliographystyle{cas-model2-names}

\bibliography{references}

\end{document}